\documentclass[a4paper,11pt]{article}
\usepackage{jheppub}
\usepackage{graphicx,amssymb,bbold,bm}
\usepackage{amsmath,amsfonts,amssymb,bm}
\usepackage{amsfonts,bm}
\usepackage{mathtools}
\usepackage[mathscr]{euscript}

\usepackage{cancel}

\usepackage{ulem}

\newcommand{\p}{\partial}
\newcommand{\pslash}{p\kern-1ex /}
\newcommand{\qslash}{q\kern-1ex /}
\newcommand{\lslash}{l\kern-1ex /}
\newcommand{\sslash}{s\kern-1ex /}
\newcommand{\kaslash}{k_a\kern-2ex /}
\newcommand{\kbslash}{k_b\kern-2ex /}
\newcommand{\Dslash}{{\cal D}\kern-1.5ex /}

\newcommand{\beqa}{\begin{eqnarray}}
\newcommand{\eeqa}{\end{eqnarray}}

\newcommand{\bpm}{\begin{pmatrix}}
\newcommand{\epm}{\end{pmatrix}}
\newcommand{\bbm}{\begin{bmatrix}}
\newcommand{\ebm}{\end{bmatrix}}

\newcommand{\bena}{\begin{eqnarray}}
\newcommand{\eena}{\end{eqnarray}}

\def\p{\partial}

\makeatletter

\@addtoreset{equation}{section}
\makeatother
\def\ben{\begin{equation}}
\def\een{\end{equation}}
\def\bena{\begin{eqnarray}}
\def\eena{\end{eqnarray}}

\newcommand{\be}{\begin{align}}
\newcommand{\ee}{\end{align}}
\newcommand{\bea}{\begin{eqnarray}}
\newcommand{\eea}{\end{eqnarray}}
\newcommand{\beas}{\begin{eqnarray*}}
\newcommand{\eeas}{\end{eqnarray*}}

\newcommand{\bdyg}{\mathcal{G}}
\newcommand{\bdyR}{\mathcal{R}}

\newcommand{\pExp}[1]{\langle~#1~\rangle}

\newcommand{\bdyD}{\mathcal{D}}

\def\scriptlap{{\kern1pt\vbox{\hrule height 0.8pt\hbox{\vrule width 0.8pt
  \hskip2pt\vbox{\vskip 4pt}\hskip 2pt\vrule width 0.4pt}\hrule height 0.4pt}
  \kern1pt}}

\def\Biggg#1{{\hbox{$\left#1\vbox to 25pt{}\right.\n@space$}}}
\def\n@space{\nulldelimiterspace=0pt \m@th}
\def\m@th{\mathsurround = 0pt}

\vspace*{0cm}

\title{The generalized entropy in holographic semiclassical de Sitter spacetime}

\medskip 

\author[a,b]{Akihiro Ishibashi,} 
\author[c]{Kengo Maeda,}
\author[d]{Takashi Okamura}

\affiliation[a]{\it Department of Physics, Nagoya University, Nagoya 464-6802, Japan}

\affiliation[b]{\it Kobayashi-Maskawa Institute, Nagoya University, Nagoya 464-8602, Japan}

\affiliation[c]{\it Faculty of Engineering, Shibaura Institute of Technology, Saitama 330-8570, Japna} 

\affiliation[d]{\it Department of Physics and Astronomy, Kwansei Gakuin University, Sanda, Hyogo, 669-1330, Japan} 

\emailAdd{ishibashi.akihiro.r7@f.mail.nagoya-u.ac.jp}\emailAdd{maeda302@sic.shibaura-it.ac.jp}
\emailAdd{tokamura@kwansei.ac.jp}

\abstract{We consider generalized de Sitter entropy which includes both entanglement entropy of the strongly coupled quantum field and the gravitational Wald entropy in the framework of holographic semiclassical gravity. The quantum corrected de Sitter spacetime is constructed perturbatively by considering static deformations of the classical solution. By applying the Noether charge formula to our semiclassical model, we show that both the renormalized entropies remain finite without introducing a cutoff, even when the bulk AdS gravitational field does not satisfy the Dirichlet boundary condition. We show that the first order variation of the generalized entropy
with respect to the de Sitter horizon always vanishes when the conformal boundary metric is promoted to a dynamical variable.}


\preprint{NU-QG-29}  

\date{} 

\begin{document}

\maketitle

\section{Introduction}\label{sec:1}
The generalized entropy, which consists of the gravitational entropy associated with a horizon and the entanglement entropy between the inside and outside of the horizon, has attracted considerable attention in a variety of research areas. The generalized second law (GSL) of black holes states that the generalized entropy never decreases toward the future, even though the gravitational entropy itself can decrease during black hole evaporation. The non-decreasing property of the generalized entropy has recently become a key framework for quantum substitutes for classical energy conditions, giving rise to concepts such as the quantum null energy condition (QNEC) and the quantum focusing conjecture (QFC)~\cite{Bousso:2015mna}. The latter states that the quantum expansion, defined as the rate of change of the generalized entropy per unit area, is non-increasing along null directions. More recently, the classical Penrose singularity theorem~\cite{Penrose} has been generalized by replacing the classical null energy condition with the GSL~\cite{Wall:2010jtc,Bousso:2025xyc,Engelhardt:2026qqs}, so that it applies even to spacetimes in which the NEC is violated. 
In this context, the quantum extremal surface~\cite{Engelhardt:2014gca}, defined as the surface that extremizes the generalized entropy along both the future and past null directions, is the quantum counterpart of a classical marginally trapped surface and is a key ingredient of the island formula~\cite{Penington:2019npb} for reproducing the Page curve~\cite{Page:1993wv} of evaporating black holes (see, e.g., a recent review \cite{Iizuka:2025xnd} for comprehensive discussions on concepts bridging classical gravity and quantum theory through the use of generalized entropy). 

Although the calculation of entanglement entropy is generally a difficult task in curved spacetimes, it becomes tractable 
for strongly coupled quantum field theories with gravitational duals through the Ryu-Takayanagi formula~\cite{Ryu:2006bv}
in the framework of the AdS/CFT duality~\cite{Maldacena:1997re}. In a static spacetime, the holographic entanglement 
entropy of a boundary region ${\cal R}$ is given by the area of the bulk minimal surface homologous to ${\cal R}$. 
In particular, when the boundary of ${\cal R}$ coincides with a horizon on the AdS boundary, the corresponding minimal 
surface is the bulk horizon. In general, the area of the minimal surface diverges near the AdS boundary, reflecting the 
UV-divergence of the boundary field theory, and a cutoff is therefore required to obtain a finite entropy. However, when the conformal boundary metric is treated as non-dynamical, i.e., the boundary metric is fixed as in the standard AdS/CFT framework, the variation of the entanglement entropy is finite and satisfies the first law of entanglement 
entropy~\cite{Faulkner:2013ica}~(see also \cite{Ishibashi:2021fcc} for the curved boundary case). 
From the bulk perspective, this follows from the Dirichlet boundary condition imposed on the metric perturbation, 
under which the variation of the bulk horizon area remains finite and can be derived using the Noether charge 
formalism~\cite{Iyer:1994ys}.

It is therefore natural to ask what happens when the boundary metric is promoted to a dynamical variable. As first proposed in Ref.~\cite{Compere:2008us}, the $d$-dimensional semiclassical Einstein equations (SCE) can be holographically encoded in mixed boundary conditions imposed at the conformal boundary of the $(d+1)$-dimensional AdS spacetime~\cite{Ishibashi:2023luz}. 
At first sight, one might expect that the variation of the entanglement entropy also becomes divergent because the bulk metric perturbations exhibit UV-divergences near the AdS boundary. To address this question, in this paper we investigate the entanglement entropy of de Sitter spacetime coupled to a strongly interacting CFT through the holographic SCE. 
Entanglement entropy in de Sitter spacetime has attracted considerable attention in both quantum field theory and holographic approaches~\cite{Maldacena:2012xp,Narayan:2015vda}. In particular, understanding quantum entanglement across the cosmological horizon is expected to shed light on the microscopic interpretation of the Gibbons--Hawking entropy~\cite{Gibbons:1977mu} and quantum aspects of de Sitter spacetime. Furthermore, recent developments have highlighted the generalized entropy as the natural quantity for incorporating quantum corrections to horizon entropy in semiclassical gravity~\cite{Chandrasekaran:2022cip}. These developments provide strong motivation for investigating the generalized entropy in semiclassical de Sitter spacetimes.

In this paper, we show that the renormalized entanglement entropy is finite without introducing a cutoff scale by applying the Noether charge formula~\cite{Iyer:1994ys} to the static perturbations of the bulk and the dynamical 
boundary metric. Furthermore, we prove that the first-order variation of the generalized de Sitter entropy vanishes for solutions of the semiclassical Einstein equations with higher-curvature corrections for $d=3,4,5$. The physical picture behind this result can be understood as follows. When the boundary spacetime is dynamical, the de Sitter horizon evolves, and hence the energy $E_{\rm CFT}$ of the CFT enclosed by the de Sitter horizon can decrease through energy transfer across the horizon. Consequently, the gravitational (Wald) entropy $S_{\rm Wald}$~\cite{Wald:1993nt} of the 
de Sitter horizon increases. For the temperature $T$, the first law takes the form
\begin{align}
\label{Bdy_first_law}
T dS_{\rm Wald}=-dE_{\rm CFT}.
\end{align}

On the other hand, there exists a bulk horizon that terminates at the de Sitter horizon on the AdS boundary. Although the variation of the horizon area itself is divergent, the Noether charge combined with the counterterm action remains finite without introducing a cutoff scale. According to the Ryu--Takayanagi formula~\cite{Ryu:2006bv}, this finite quantity gives the entanglement entropy $S_{\rm EE}$ of the region enclosed by the de Sitter horizon. Applying the Noether charge formula~\cite{Iyer:1994ys}, as in the case of a non-dynamical boundary metric~\cite{Faulkner:2013ica}, we obtain the first law of entanglement entropy,
\begin{align}
\label{Bulk_first_law}
T dS_{\rm EE}=dE_{\rm CFT}.
\end{align}
Combining Eqs.~(\ref{Bdy_first_law}) and (\ref{Bulk_first_law}), we find
\begin{align}
\label{GSL}
d(S_{\rm EE}+S_{\rm Wald})=0.
\end{align}
Although our derivation is based on static perturbations, we expect that the same relation holds for dynamical processes. 
Therefore, Eq.~(\ref{GSL}) can be regarded as a saturated form of the generalized second law, where the total 
generalized entropy remains unchanged. 

The paper is organized as follows. In Section 2, we briefly review the holographic semiclassical model of de Sitter spacetime~\cite{Ishibashi:2025gvl}. In Section 3, we show that the sum of the variations of the bulk charge 
associated with the entanglement entropy and the boundary gravitational charge associated with the Wald entropy 
vanishes, confirming Eq.~(\ref{GSL}). In Section 4, we calculate the static perturbations of de Sitter spacetime and 
show that each renormalized charge is finite and that Eq.~(\ref{GSL}) is indeed satisfied. 
Section 5 is devoted to a summary and discussion. In the appendix, we collect some relevant formulas.

\section{The holographic semiclassical model of de Sitter spacetime}\label{sec:2}
We begin with a brief review of a holographic model of $d$-dimensional de Sitter spacetimes dynamically interacting with a strongly coupled quantum field via the the semiclassical Einstein equations with higher curvature 
corrections~\cite{Ishibashi:2025gvl}. We write the $d+1$-dimensional bulk metric as 
\begin{align}
\label{bulk_metric}
ds_{d+1}^2 &=G_{MN}dX^MdX^N \nonumber \\
&=\frac{dz^2}{\Omega^2(z)}+g_{\mu\nu}(x,z)dx^\mu dx^\nu \nonumber \\
&=\frac{dz^2+\tilde{g}_{\mu\nu}(x,z)dx^\mu dx^\nu}{\Omega^2(z)}, \qquad 
\Omega(z)=\frac{\ell}{L}\sinh\left(\frac{z}{\ell}  \right),  
\end{align}
where $L$ and $\ell$ are the bulk AdS and the boundary curvature radius, respectively. 
We define the conformal AdS boundary metric $\bdyg_{\mu\nu}(x)$ by 
\begin{align}
\label{conformal_metric}
\bdyg_{\mu\nu}(x)=\lim_{z\to 0}\tilde{g}_{\mu\nu}(x,z) .
\end{align}  
The conformal metric satisfies the semiclassical Einstein~(SCE) equations with 
higher curvature corrections,  
\begin{subequations}
\label{semi_Eqs} 
\begin{align}
  & {\cal E}_{\mu\nu} = 8\, \pi\, G_{d}\, \pExp{{\cal T}_{\mu\nu}}, 
\label{semi_tensor} \\
  & {\cal E}_{\mu\nu}
  := \bdyR_{\mu\nu}-\frac{\bdyR}{2}\bdyg_{\mu\nu}+\Lambda_{d}\,\bdyg_{\mu\nu}
  + \alpha_1(\mu) \mathcal{H}^{(1)}_{\mu\nu}
  +\alpha_2(\mu) \mathcal{H}^{(2)}_{\mu\nu}+\alpha_3(\mu) \mathcal{H}^{(3)}_{\mu\nu}, 
\label{eq:def-calE}
\end{align}
\end{subequations}
where $G_d$ is the $d$-dimensional gravitational constant at the AdS boundary, $z=0$, 
${\bdyR}_{\mu\nu}$ and $\mathcal{H}^{(i)}_{\mu\nu}$~($i=1,2,3$) are the Ricci tensor and the higher curvature tensors respectively, constructed from the metric $\bdyg_{\mu\nu}$ as defined below~(\ref{def:H}). In general, the coefficients $\alpha_i(\mu)~(i=1,2,3)$ are parameters that depend on an 
arbitrary mass scale associated with the renormalization of the field theory. In what follows, 
we suppress the explicit \(\mu\)-dependence of these coefficients for notational simplicity. 
Finally, $\pExp{{\cal T}_{\mu\nu}}$ represents the vacuum expectation value of the stress-energy tensor of 
the strongly interacting quantum field with a gravitational dual. 

As the boundary gravity action, we adopt the quadratic gravity model with the action 
\begin{align}
\label{bdy_action}
S_\text{bdy}
 &= \int \sqrt{-\bdyg}L_\text{bdy}\,d^d x \nonumber \\
 &=\frac{1}{16\pi G_d}\int \sqrt{-\bdyg}\left(\bdyR-2\Lambda_d+\alpha_1 \bdyR^2
+\alpha_2\bdyR_{\mu\nu} \bdyR^{\mu\nu}+\alpha_3 \bdyR_{\mu\nu\alpha\beta}\bdyR^{\mu\nu\alpha\beta}\right)d^dx, 
\end{align}
where $\bdyR_{\mu\nu\alpha\beta}$ is the Riemann tensor of the metric $\bdyg_{\mu\nu}$, 
and $\Lambda_d$ is the (renormalized) cosmological constant. 
Hereafter, we suppress the volume element in integrals of scalar 
quantities, writing, for example, $S_{\text{bdy}}=\int_{\Sigma_0} L_{\text{bdy}}$ in 
Eq.~(\ref{bdy_action}), where $\Sigma_0$ is the $d$-dimensional timelike 
hypersurface at $z=0$.
By varying the action~(\ref{bdy_action}) with respect to $\bdyg_{\mu\nu}$,  
we have $\mathcal{H}^{(i)}_{\mu\nu}~(i=1,2,3)$ as  
\begin{subequations}
\label{def:H}
\begin{align}
\label{def:H1}
 \mathcal{H}^{(1)}_{\mu\nu}=2(\bdyR_{\mu\nu}-\bdyD_\mu\bdyD_\nu)\bdyR-
\bdyg_{\mu\nu}\left(\frac{1}{2}\bdyR^2-2 \bdyD^2 \bdyR   \right), 
\end{align}
\begin{align}
\label{def:H2}
\mathcal{H}^{(2)}_{\mu\nu} =2\bdyR_{\mu\rho\nu\sigma}\bdyR^{\rho\sigma}
+\bdyD^2 \bdyR_{\mu\nu}
-\bdyD_\mu\bdyD_\nu\bdyR
-\frac{1}{2}\bdyg_{\mu\nu}(\bdyR_{\sigma\rho}\bdyR^{\sigma\rho}- \bdyD^2\bdyR), 
\end{align}
\begin{align}
\label{def:H3}
\mathcal{H}^{(3)}_{\mu\nu}=2\bdyR_{\mu\rho\sigma\tau}{\bdyR_\nu}^{\rho\sigma\tau}
- \frac{\bdyg_{\mu\nu}}{2}\, \bdyR_{\alpha\beta}^{\rho\sigma}\,
  \bdyR^{\alpha\beta}_{\rho\sigma}
+4{\bdyR}_{\mu\rho\nu\sigma}\bdyR^{\rho\sigma}
-4{\bdyR}_{\mu\rho}{\bdyR^{\rho}}_\nu
-2\bdyD_\mu\bdyD_\nu \bdyR+4 \bdyD^2 \bdyR_{\mu\nu}, 
\end{align}
\end{subequations}
where $\bdyD_\mu$ is the covariant derivative with respect to $\bdyg_{\mu\nu}$. 

According to the AdS/CFT duality~\cite{Maldacena:1997re}, the vacuum 
expectation value of the stress-energy tensor $\pExp{{\cal T}_{\mu\nu}}$ is 
obtained by varying the renormalized on-shell bulk action given by  
\begin{eqnarray}
\label{bulk_action}
\Gamma 
 &=& \int_M L+\int_{\Sigma_\varepsilon} \frac{K}{8\pi G_{d+1}}
+\int_{\Sigma_\varepsilon} L_\text{ct}, \nonumber \\
  L &=&\frac{1}{16\pi G_{d+1}}\left(R+\frac{d(d-1)}{L^2}\right),  \nonumber \\
  L_\text{ct} \! &:=& \! -\frac{L}{16\pi G_{d+1}(d-2)} \! 
\left( \! 
      \frac{2(d-1)(d-2)}{L^2} \! + R(g) \! + \! \frac{c_dL^2}{d-2} \!
\left\{ \! R_{\mu\nu}(g)R^{\mu\nu}(g) \! - \! \frac{d\,R^2(g)}{4(d-1)} \right\} \!
\right).  
\end{eqnarray} 
Here $\Sigma_\varepsilon$ denotes the timelike hypersurface at $z=\varepsilon$, and the associated extrinsic curvature 
$K_{\mu\nu}$ is defined by 
\begin{align}
\label{ext_curvature}
K_{\mu\nu}:=-\frac{1}{2}\Omega \partial_zg_{\mu\nu}, 
\end{align}
and $c_d$ is given by 
\begin{eqnarray} 
  c_d &=& \left\{
\begin{array}{lll}
 0 & \quad d = 3 \\
 - \ln\left(\mu z  \right) & \quad d=4 \\
 1 & \quad d=5 
\label{bulkct} 
\end{array}
\right.,  
\end{eqnarray}
where $\mu$ is an arbitrary mass scale related to renormalization in the field theory. 

As shown in Refs.~\cite{Compere:2008us, Ishibashi:2023luz}, by varying the total action 
$S_\text{total}= \Gamma+S_\text{bdy}$ with respect to the metric $G_{MN}$ and $\bdyg_{\mu\nu}$ 
simultaneously, we derive the semiclassical equations~(\ref{semi_Eqs}) and the vacuum Einstein equations, whose solutions are realized by allowing a mixed boundary condition on the AdS boundary, instead of the Dirichlet boundary condition. As shown in Ref.~\cite{Ishibashi:2025gvl}, the SCE Eqs.~(\ref{semi_Eqs}) are satisfied for the bulk spacetime~(\ref{bulk_metric}) with the $d$-dimensional de Sitter metric
\begin{align} 
\label{BG_dS_metric}
 &\overline{\bdyg}_{\mu\nu}dx^\mu dx^\nu 
  \left( = {\Omega^2} \overline{g}_{\mu\nu}dx^\mu dx^\nu \right) 
 = -\frac{f(u)}{u}dt^2+\frac{\ell^2}{4u^2f(u)}du^2+\frac{\ell^2}{u}d\Omega_{d-2}^2 , 
\end{align}
where $f(u)=u-1$ for a suitable choice of the cosmological conatant, $\Lambda_d$, and 
where $d\Omega_{d-2}^2$ denotes the metric of the unit $d-2$-dimensional sphere. 
This de Sitter metric in the static chart has the cosmological Killing horizon at $u=1$ with $(d-2)$-dimensional sphere as the bifurcate surface, which divides each de Sitter slice $(\Sigma_\varepsilon, \overline{\bdyg}_{\mu\nu})$ into two static wedges just like the Schwarzschild spacetime.  
 
\begin{figure}[htbp]
  \begin{center}
  \includegraphics[width=150mm]{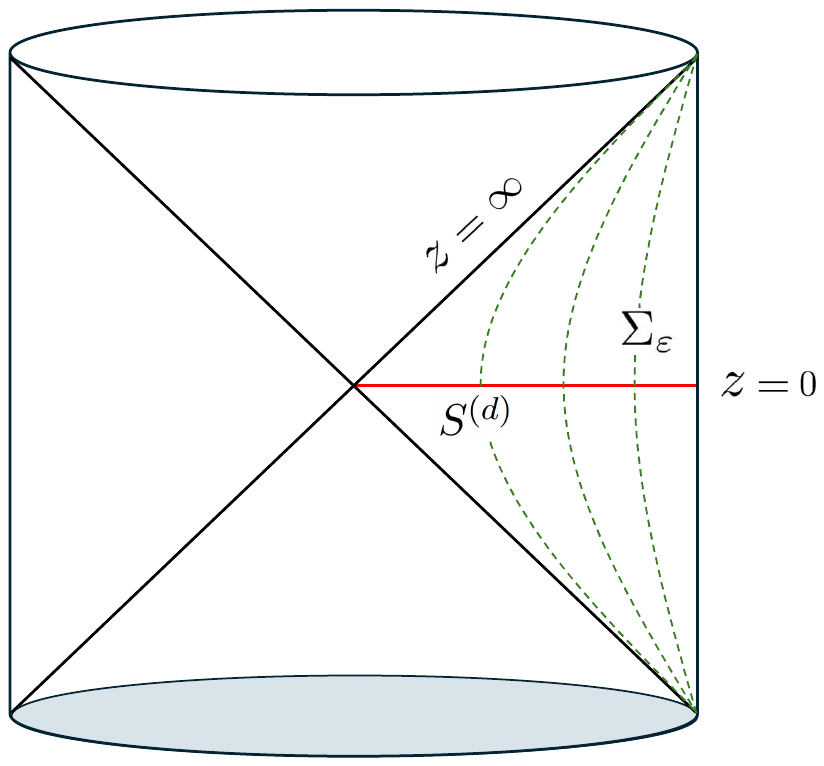}
  \end{center}
  \caption{\small AdS bulk spacetime foliated by de Sitter slices. $z=\infty$ corresponds to the spherial Rindler horizon, whose inside is expressed as the hyperbolic FLRW chart of AdS spacetime.  
 }
  \label{dS_foliation}
 \end{figure} 

\begin{figure}[htbp]
  \begin{center}
  \includegraphics[width=180mm]{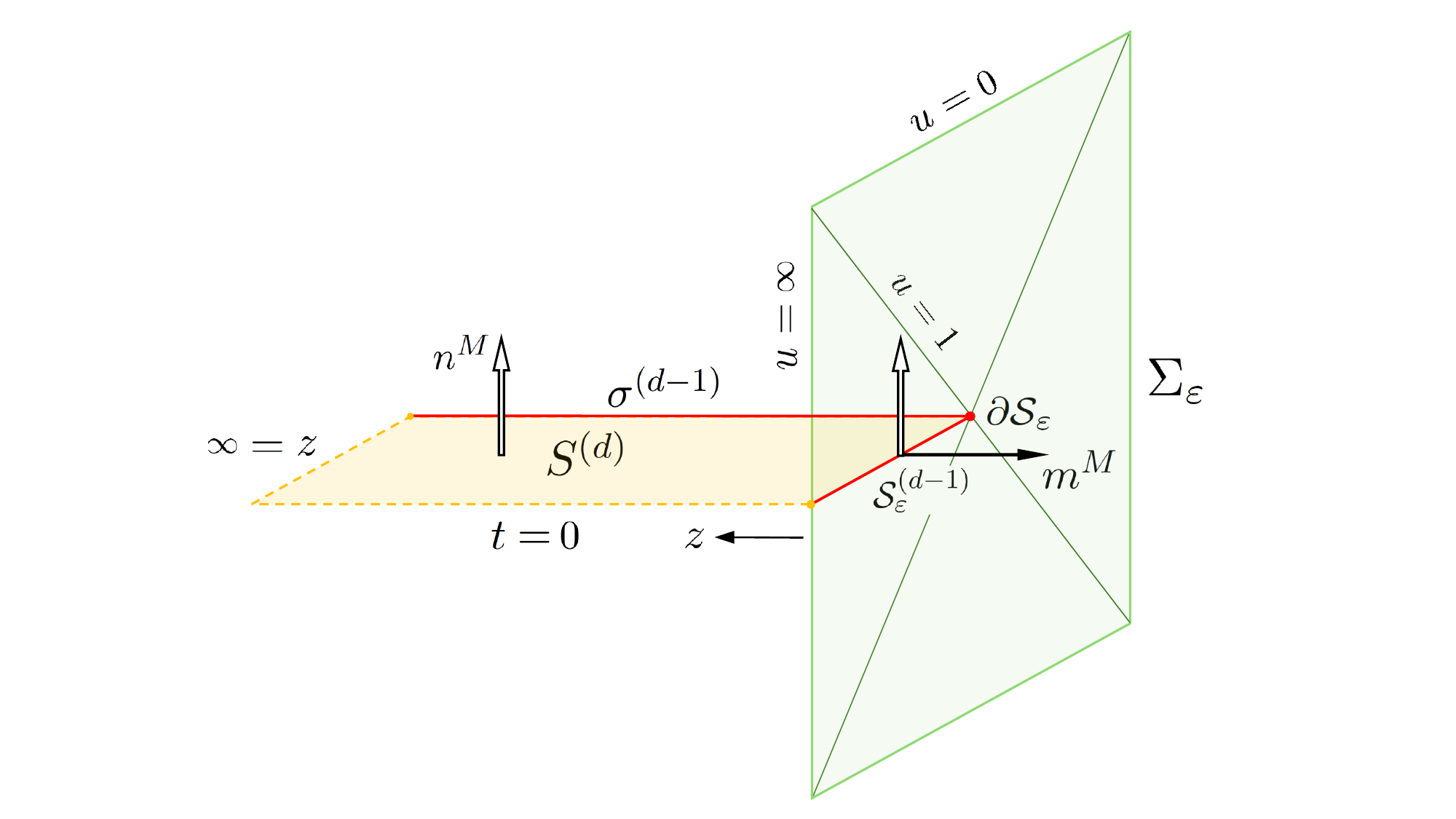}
  \end{center}
  \caption{\small 
The surface $S^{(d)}$ defined as a part of $t=0$ hypersurface and its boundaries ${\cal S}^{(d-1)}_\varepsilon$, $\sigma^{(d-1)}$ and $\Sigma_\varepsilon$. The vectors $n^M$ and $m^M$ are introduced later.    
 }
  \label{SSsigma}
 \end{figure}

The AdS-bulk metric is now concretely written in terms of the above de Sitter metric as 
\begin{eqnarray}
ds^2_{(d+1)} &=& \dfrac{1}{\Omega^2} \left(dz^2+ \overline{\bdyg}_{\mu\nu}dx^\mu dx^\nu \right) 
\nonumber \\ 
 &=& \dfrac{L^2}{\ell^2 \sinh^2(z/\ell)} \left[ dz^2 -\frac{f(u)}{u}dt^2+\frac{\ell^2}{4u^2f(u)}du^2+\frac{\ell^2}{u}d\Omega_{d-2}^2 \right] , 
\end{eqnarray}
whose geometry is illustrated in Fig.~\ref{dS_foliation}, with the bulk spherical Rindler horizon located at $z=\infty$ and the SCE satisfied for the conformal metric $\bdyg_{\mu\nu}$ at AdS boundary (at $z=0$). 
In terms of this coordinate system, we introduce several geometric quantities (see Fig.~\ref{SSsigma}). 
First, let ${\cal S}^{(d-1)}_\varepsilon$ be the $(d-1)$-dimensional spacelike surface on 
$\Sigma_\varepsilon$, specified by $t=0$ and $1<u<\infty$ in one of the static wedges (say, the left wedge) of de Sitter slice $(\Sigma_\varepsilon, \overline{\bdyg}_{\mu\nu})$, so that it is topologically ${\cal S}_\varepsilon^{(d-1)} \simeq{\Bbb S}^{d-2}\times {\Bbb R}$ and its boundary $\partial {\cal S}_\varepsilon^{(d-1)}$ corresponds to the bifurcate surface of the de Sitter horizon on $\Sigma_\varepsilon$. 
Next, let $\sigma^{(d-1)}$ be the $(d-1)$-dimensional spacelike surface in the bulk, specified by $t=0$, $u=1$, and $\varepsilon < z<\infty$ so that its boundary at $z=\varepsilon$ coincides with the bifurcate surface of the de Sitter horizon on $\Sigma_\varepsilon$, hence expressing the ``bifurcate surface string" with topology ${\sigma}^{(d-1)} \simeq {\Bbb S}^{d-2}\times {\Bbb R}$. 
Finally, let $S^{(d)}$ be the $d$-dimensional spacelike submanifold in the bulk, specfied by $t=0$, $1<u < \infty$, and $\varepsilon < z<\infty$, so that it is bounded by ${\cal S}^{(d-1)}_\varepsilon$ 
and $\sigma^{(d-1)}$
(with taking the limit $\varepsilon \mapsto 0$) and the vertex ($t=0$ and $z \rightarrow \infty$) of the bulk spherical Rindler horizon.

\section{The first law of generalized entropy}\label{sec:3}
Let us consider static perturbations of the AdS bulk spacetime~(\ref{bulk_metric}) with the background 
de Sitter spacetime~(\ref{BG_dS_metric}) on the conformal AdS boundary. 
We assume that the perturbed metric takes the form 
\begin{align}
\label{pertb_metric}
g_{\mu\nu}(x,z)=\frac{\overline{\bdyg}_{\mu\nu}(x)+\epsilon h_{\mu\nu}(x,z)}{\Omega^2(z)}
\end{align}
and impose the transverse-traceless conditions 
\begin{align}
\label{TT_cond}
{h^\mu}_\mu=\overline{\bdyD}^\nu h_{\mu\nu}=0, 
\end{align}
where $\overline{\bdyD}^\nu$ is the covariant derivative with respect to the background conformal 
boundary metric $\overline{\bdyg}_{\mu\nu}(x)$. 
As shown below, we derive Eq.~(\ref{GSL}) with respect to the generalized entropy by applying the Noether charge formula~\cite{Iyer:1994ys} to both the bulk and the dynamical boundary spacetimes. 
In what follows, boldface letters denote differential forms with their indices indicating the corresponding components. 
Accordingly, the Lagrangian $L$ in an $n$-dimensional spacetime can be expressed as an $n$-form,
${\bm L}=L {\bm\epsilon}_{a_1a_2\cdots a_n}$, where ${\bm\epsilon}_{a_1a_2\cdots a_n}$ 
denotes the Levi-Civita volume form. 
As we already used in the previous section, throughout this paper, we use the indices $M,\,N,\,\cdots$ to denote bulk spacetime coordinates, while the indices $\mu,\,\nu,\,\cdots$ to denote coordinates on the (conformal) boundary spacetime or on a constant-$z$ hypersurface, as appropriate.
When there is no risk of confusion, we often omit the indices. In such cases, however, we explicitly indicate the degree of the differential form. For instance, the $d$-dimensional Levi-Civita volume form ${\bm\epsilon}_{\mu_1\mu_2\cdots\mu_d}$ is simply written as ${\bm\epsilon}^{(d)}$. Also in the following, the tilde indicates that the corresponding quantity is defined with respect to the conformal metric $\bdyg_{\mu\nu}$~(\ref{conformal_metric}). 
In particular, ${\bm\epsilon}^{(d)}$ and ${ \tilde{\bm\epsilon}}{}^{(d)}$ 
denote the $d$-dimentional Levi-Civita volume forms associated with $g_{\mu\nu}$ and 
$\bdyg_{\mu\nu}=\lim_{z\to 0}\Omega^2g_{\mu\nu}$, respectively.

The variation of the Lagrangian $L$, $L_\text{ct}$ in (\ref{bulk_action}), and $L_\text{bdy}$ in 
(\ref{bdy_action}) can be expressed in the form 
\begin{align} 
\label{variation_Lg}
& \delta {\bm L} =({\bm E})^{MN}\delta G_{MN} + d{\bm \Theta}^{(d)} , 
\nonumber \\
& \delta {\bm L}_\text{ct}=({\bm E}_\text{ct})^{\mu\nu}\delta g_{\mu\nu}+d{\bm \Theta}_\text{ct}^{(d-1)} , 
\nonumber \\
& \delta \tilde{{\bm L}}_\text{bdy}=(\tilde{{\bm E}}_\text{bdy})^{\mu\nu}\delta \bdyg_{\mu\nu}+
d \tilde{{\bm \Theta}}_\text{bdy}^{(d-1)} , 
\end{align}
where $({\bm E})^{MN}:={\bm \epsilon}^{(d+1)}E^{MN}$, 
$({\bm E}_\text{ct})^{\mu\nu}:={\bm \epsilon}^{(d)}E_\text{ct}^{\mu\nu}$, and 
$(\tilde{{\bm E}}_\text{bdy})^{\mu\nu}:={\tilde{\bm \epsilon}}^{(d)}E_\text{bdy}^{\mu\nu}$,   
with the tensors $E^{MN}$, $E_\text{ct}^{\mu\nu}$, and $E_\text{bdy}^{\mu\nu}$, given by 
\begin{subequations}
\label{Eij}
\begin{align}
\label{Eij_bulk}
& E^{MN}=\frac{1}{16\pi G_{d+1}}\left(-R^{MN}
+\frac{1}{2}RG^{MN}+\frac{d(d-1)}{2L^2}G^{MN} \right),
\end{align}
\begin{align}
\label{Eij_ct}
& E_\text{ct}^{\mu\nu}=
-\frac{1}{16\pi G_{d+1}}
\Biggl[\frac{d-1}{L}g^{\mu\nu}-\frac{L}{d-2}
\left(R^{\mu\nu}-\frac{1}{2}Rg^{\mu\nu}\right)
-\frac{c_dL^3}{(d-2)^2}\left(H^{\mu\nu}_{(2)}-\frac{d}{4(d-1)}H^{\mu\nu}_{(1)} \right)\Biggr], 
\end{align}
\begin{align}
\label{Eij_bdy}
& E_\text{bdy}^{\mu\nu}=-\frac{1}{16\pi G_d}{\cal E}^{\mu\nu},
\end{align}
\end{subequations}
while ${\bm \Theta}^{(d)}$, ${\bm \Theta}_\text{ct}^{(d-1)}$, and $\tilde{\bm \Theta}_\text{bdy}^{(d-1)}$ denote the corresponding symplectic potential forms introduced in Eqs.~(\ref{symplectic_form}) in Appendix~\ref{Appendix_Symplectic}. 
Here, $H^{\mu\nu}_{(1)}$ and $H^{\mu\nu}_{(2)}$ are defined by the same expressions as $\mathcal{H}^{\mu\nu}_{(1)}$ 
and $\mathcal{H}^{\mu\nu}_{(2)}$, respectively, with the boundary metric $\bdyg_{\mu\nu}$ replaced by $g_{\mu\nu}$.
Note that the indices of $E_{\text{ct}}^{\mu\nu}$ and $E_{\text{bdy}}^{\mu\nu}$ are 
raised and lowered using $g_{\mu\nu}$ and $\bdyg_{\mu\nu}$, respectively.

In terms of an arbitrary vector $\xi^M$ tangent to $z=\text{const.}$ hypersurface, let us define the bulk and 
the boundary Noether currents, ${\bm J}^{(d)}$ and ${\bm J}^{(d-1)}:= {\bm J}_\text{ct}^{(d-1)} +\tilde{{\bm J}}_\text{bdy}^{(d-1)}$, by 
\begin{eqnarray}
\label{def:NC} 
{\bm J}^{(d)} &=& {\bm \Theta}^{(d)}(G, {\cal L}_\xi G) - L i_\xi  {\bm\epsilon}^{(d+1)} , 
\nonumber \\
 {\bm J}_\text{ct}^{(d-1)} &=& {\bm \Theta}_\text{ct}^{(d-1)}(g, {\cal L}_\xi g)- 
\left(L_\text{ct}+\frac{K}{8\pi G_{d+1}} \right) i_\xi  {\bm\epsilon}^{(d)} ,
\nonumber \\
\tilde{{\bm J}}_\text{bdy}^{(d-1)} 
   &=& \tilde{{\bm \Theta}}_\text{bdy}^{(d-1)}(\bdyg, {\cal L}_\xi \bdyg)
-L_\text{bdy} i_\xi \tilde{\bm\epsilon}^{(d)}, 
\end{eqnarray}
where $K={K^\mu}_\mu$ and here and in the following $i_\xi$ denotes the interior product with $\xi$.

Let us introduce the bulk Noether charge $(d-1)$-form as, 
\begin{align}
\label{bulk_NC_form}
 {\bm Q}^{(d-1)}
 = \frac{1}{2}{\bm \epsilon}_{MNK_3K_4\cdots K_{d+1}}Q^{MN}, \qquad 
Q^{MN}:=\frac{1}{8\pi G_{d+1}}\nabla^{[N} \xi^{M]}. 
\end{align} 
Using the relation
${\bm J}^{(d)}
= d{\bm Q}^{(d-1)}(\xi)
-2{\bm \epsilon}_{M K_1 \cdots K_{d+1}}{E^M}_N\, \xi^N$, we can express the integral of
${\bm J}^{(d)}$ over the $d$-dimensional spacelike 
hypersurface $S^{(d)}$ in terms of the bulk Noether charges
$Q_\sigma(\xi)$ on $\sigma^{(d-1)}$ and $Q_E(\xi)$ on ${\cal S}^{(d-1)}_\varepsilon$ as
\begin{eqnarray} 
\label{bulk_NC}
\int_S {\bm J}^{(d)} &=& 
-2\int_S n_M E^{MN}\,\xi_N +\int_S d{\bm Q}^{(d-1)}(\xi)  
\nonumber \\
&=& \int_\sigma {\bm Q}^{(d-1)}
+\int_{{\cal S}_\varepsilon}{\bm Q}^{(d-1)}-2\int_S n_M E^{MN}\,\xi_N
\nonumber \\
&=&Q_\sigma (\xi)+Q_\text{E}(\xi).  
\end{eqnarray}
Here and in the following, $n^M$ denotes the future-directed unit timelike vector normal to 
$S^{(d)}$. 
In the last equality, we used the bulk equations of motion, $E^{MN}=0$, and defined $Q_\sigma$ and $Q_\text{E}$ accordingly. Here and hereafter, we omit the volume element in integral formulas, for notational simplicity.  

Similarly, the integration of the boundary current ${\bm J}^{(d-1)}$ over $(d-1)$-surface 
${\cal S}_{\varepsilon}^{(d-1)}$ yields 
\begin{eqnarray}
\label{bdy_NC}
 {Q}_{{\cal S}_\varepsilon} &:=& \int_{{\cal S}_\varepsilon}{\bm J}^{(d-1)}
 = \int_{{\cal S}_\varepsilon}\left\{{\bm \Theta}_\text{ct}^{(d-1)}
  - \left(L_\text{ct}+\frac{K}{8\pi G_{d+1}} \right) i_\xi  {\bm\epsilon}^{(d)} 
+\tilde{{\bm \Theta}}_\text{bdy}^{(d-1)}
-L_\text{bdy} i_\xi \tilde{\bm\epsilon}^{(d)} 
  \right\} 
\nonumber \\
&=&
\int_{{\cal S}_\varepsilon} d{\bm Q}_\text{ct}^{(d-2)}(\xi)
  - 2\int_{{\cal S}_\varepsilon} n_\mu E_\text{ct}{}^{\mu}{}_{\nu}\, \xi^\nu
- \int_{{\cal S}_\varepsilon}\frac{K i_\xi  {\bm\epsilon}^{(d)}}{8\pi G_{d+1}}
+ \int_{{\cal S}_\varepsilon} d\tilde{\bm Q}_\text{bdy}^{(d-2)}(\xi)
  - 2\int_{{\cal S}_\varepsilon} \Tilde{n}_\mu E_\text{bdy}{}^{\mu}{}_{\nu}\,
    \xi^\nu
\nonumber \\
&=&
\int_{\p {\cal S}_\varepsilon} \!\! {\bm Q}_\text{ct}^{(d-2)}(\xi)
+ \int_{\p {\cal S}_\varepsilon} \!\! \tilde{\bm Q}_\text{bdy}^{(d-2)}(\xi)
- 2\!\int_{{\cal S}_\varepsilon} \!\!n_\mu
  E_\text{ct}{}^{\mu}{}_{\nu}\, \xi^\nu \nonumber \\
&&- \int_{{\cal S}_\varepsilon} \frac{K i_\xi {\bm\epsilon}^{(d)}}{8\pi G_{d+1}}
- 2 \int_{{\cal S}_\varepsilon} \!\!\Tilde{n}_\mu
  E_\text{bdy}{}^{\mu}{}_{\nu}\, \xi^\nu
\end{eqnarray}
with the help of Eqs.~(\ref{Theta_bdy}) and (\ref{Theta_ct}), where here and hereafter $n^\mu$ denotes the future-directed unit timelike vector normal to ${\cal S}_\varepsilon$, which coincides with the pull-back to 
$\Sigma_\varepsilon$ of $n^M$ introduced above\footnote{
This is true because we define the hypersurfaces $S^{(d)}$ and $\Sigma_\varepsilon$ in terms of the coordinates of $t=const.$ and $z=const$, respectively. It is possible to define these hypersurfaces in a geometric way, so that $S^{(d)}$ can be part of a general Cauchy surface and $\Sigma_\varepsilon$ a general foliation in accord with the Fefferman-Graham expansion formula.}. 
The vector $\tilde{n}^\mu$ is the future-directed unit timelike vector normal to 
${\cal S}_\varepsilon$ with respect to the conformal metric $\bdyg_{\mu\nu}$ on the AdS boundary. 
It is related to $n^\mu$ by
${\tilde n}^\mu=n^\mu/\Omega$ and 
${\tilde n}_\mu=\bdyg_{\mu\nu}\,\tilde n^\nu=\Omega\, n_\mu$. 
Here, $\p {\cal S}_\varepsilon$ denotes the bifurcation surface of the de Sitter horizon embedded in the AdS boundary (see Fig.~\ref{SSsigma}).   
In contrast to the bulk case, the boundary current cannot be entirely rewritten in terms of boundary charges, since both $E_\text{ct}^{\mu\nu}$ and $E_\text{bdy}^{\mu\nu}$ are non-vanishing. However, as shown below, the sum of the bulk current (\ref{bulk_NC}) and the boundary current (\ref{bdy_NC}) can be reorganized into the Noether charge on the bulk horizon and the Wald entropy on the de Sitter horizon of the boundary spacetime, respectively, by virtue of the SCE Eqs.~(\ref{semi_Eqs}).

Let us define the renormalized bulk charges $Q^\text{(ren)}_\sigma$ on $\sigma^{(d-1)}$ and 
$Q^\text{(ren)}_\text{E}$ on $\Sigma_\varepsilon$ as follows: 
\begin{eqnarray} 
\label{def:ren_charge_H}
Q^\text{(ren)}_{\sigma}(\xi)
 := \lim_{\varepsilon\to 0}\left[
\int_\sigma {\bm Q}^{(d-1)}(\xi)+\int_{\p {\cal S}_\varepsilon}{\bm Q}_\text{ct}^{(d-2)}(\xi)\right], 
\end{eqnarray}
\begin{align} 
\label{def:ren_charge_E}
& Q^\text{(ren)}_\text{E}(\xi):=
\lim_{\varepsilon\to 0}\left[\int_{{\cal S}_\varepsilon}
\left({\bm Q}^{(d-1)}(\xi)
-\frac{K i_\xi  {\bm\epsilon}^{(d)}}{8\pi G_{d+1}}\right)
-\int_{{\cal S}_\varepsilon}2 n_\mu {{E_\text{ct}}^\mu}_\nu\xi^\nu \right]
=-\int_{{\cal S}_\varepsilon} \tilde{n}_\mu \pExp{ {\cal {\bm T}^\mu}_\nu }\xi^\nu, 
\end{align}
where we used Eq.~(\ref{Q_E_formula}) in Appendix~\ref{Appendix_derivation}, and where the volume element of ${\cal S}_\varepsilon^{(d-1)}$ is understood. This $Q_\sigma^{\rm (ren)}$ corresponds to the $S_\text{EE}$ in eq.~(\refeq{Bulk_first_law}).  
In the next section, we show that $Q^\text{(ren)}_\sigma (\xi)$ remains finite in the limit $z=\varepsilon\to 0$. 
Thus, combining Eqs.~(\ref{bulk_NC}), (\ref{bdy_NC}), (\ref{def:ren_charge_H}), and (\ref{def:ren_charge_E}), 
we obtain 
\begin{eqnarray} 
\label{J+j}
\lim_{\varepsilon\to 0}(Q_\sigma +Q_\text{E} + {Q}_{{\cal S}_\varepsilon} )
 &=& \lim_{\varepsilon\to 0}
\Bigg[ 
\int_\sigma \!\! {\bm Q}^{(d-1)}(\xi) + \int_{\p {\cal S}_\varepsilon} \!\! {\bm Q}_\text{ct}^{(d-2)}(\xi)+
\int_{{\cal S}_\varepsilon} \!\!  {\bm Q}^{(d-1)}(\xi)
-\int_{{\cal S}_\varepsilon}\frac{K \, i_\xi  {\bm\epsilon}^{(d)}}{8\pi G_{d+1}} 
\nonumber \\
&{}& \qquad \quad 
-2\int_{{\cal S}_\varepsilon} \! n_\mu  E_\text{ct}{}^{\mu}{}_{\nu} \xi^\nu
-2\int_{{\cal S}_\varepsilon} \! \tilde{n}_\mu E_\text{bdy}{}^{\mu}{}_{\nu} \xi^\nu
+\int_{\p {\cal S}_\varepsilon}\!\! \tilde{\bm Q}_\text{bdy}^{(d-2)}(\xi)\Bigg] \nonumber \\
&=& Q^\text{(ren)}_\sigma(\xi)-\int_{{\cal S}_\varepsilon} \! \tilde{n}_\mu \pExp{ {{\cal T}^\mu}_\nu }\xi^\nu
-2\int_{{\cal S}_\varepsilon} \! \tilde{n}_\mu E_\text{bdy}{}^{\mu}{}_{\nu} \xi^\nu
+\int_{\p {\cal S}_\varepsilon}\tilde{\bm Q}_\text{bdy}^{(d-2)}(\xi) \nonumber \\
&=& Q^\text{(ren)}_\sigma(\xi)+\int_{\p {\cal S}_\varepsilon} \! \tilde{\bm Q}_\text{bdy}^{(d-2)}(\xi), 
\end{eqnarray}
where we used the SCE Eqs.~(\ref{semi_Eqs}) in the last equality. This is the generalized entropy as the sum of entanglement entropy and gravitational one appeared in (\ref{GSL}). This result shows that the sum of the bulk and boundary currents can be expressed entirely in terms of 
Noether charges, although the boundary current itself cannot be entirely rewritten in terms of boundary charges. 
This feature is a direct consequence of imposing the semiclassical Einstein equations on the boundary metric.

Now, let us take the first-order variation of Eq.~(\ref{J+j}) under the static perturbations~(\ref{pertb_metric}) satisfying 
Eq.~(\ref{TT_cond}). Substituting Eqs.~(\ref{variation_Lg}) into the variations of ${\bm J}^{(d)}$ and ${\bm J}^{(d-1)}$ 
in Eqs.~(\ref{def:NC}) and applying the formula for an aribitrary form ${\bm u}$, 
\begin{align}
\label{formula_form}
{\cal L}_\xi {\bm u}=\xi\cdot d{\bm u}+d(\xi\cdot {\bm u}), 
\end{align}
one obtains 
\begin{eqnarray} 
\label{variation_J+j}
\delta(Q_\sigma + Q_\text{E} + {Q}_{{\cal S}_\varepsilon})
&=&\int_S (\delta {\bm \Theta}^{(d)}(\overline{G},\,{\cal L}_\xi G)
- {\cal L}_\xi{\bm \Theta}^{(d)}(\overline{G},\,\delta G)) 
+\int_\sigma i_\xi  {\bm \Theta}^{(d)}(\overline{G},\,\delta G) 
\nonumber \\
&{}& -\frac{1}{8\pi G_{d+1}}\int_{{\cal S}_\varepsilon}
   \delta(K(\overline{g},\,\delta g) i_\xi  {\bm\epsilon}^{(d)}) 
\nonumber \\
&{}&
+\int_{{\cal S}_\varepsilon}
                i_\xi 
                     \left\{ 
                              {\bm \Theta}^{(d)}(\overline{G},\,\delta G)
                            - ({\bm E}_\text{ct})^{\mu\nu}(\overline{g}) \delta g_{\mu\nu}
                            - (\tilde{\bm E}_\text{bdy})^{\mu\nu}(\overline{\bdyg})\delta\bdyg_{\mu\nu} 
                     \right\} 
\nonumber \\
&{}& +\int_{\p {\cal S}_\varepsilon} 
                     i_\xi  \left\{
                                     {\bm \Theta}_\text{ct}^{(d-1)}(\overline{g},\,\delta g) 
                                     + \tilde{\bm \Theta}_\text{bdy}^{(d-1)}(\overline{\bdyg},\,\delta\bdyg)
                            \right\} 
\nonumber \\ 
&{}& \hspace{-40mm}
+\int_{{\cal S}_\varepsilon} \{\delta {\bm \Theta}_\text{ct}^{(d-1)}(\overline{g},\,{\cal L}_\xi g)
-{\cal L}_\xi{\bm \Theta}_\text{ct}^{(d-1)}(\overline{g},\,\delta g)
+\delta\tilde{\bm \Theta}_\text{bdy}^{(d-1)}(\overline{\bdyg},\,{\cal L}_\xi \bdyg)
-{\cal L}_{\xi}\tilde{\bm \Theta}_\text{bdy}^{(d-1)}(\overline{\bdyg},\,\delta \bdyg) \}, 
\end{eqnarray}
where we used the bulk equations of motion, ${\bm E}^{MN}=0$. 

Eq.~(\ref{variation_J+j}) can be further simplified. First, we note that    
${\bm \Theta}^{(d)}(\overline{G},\,\delta G)$ and $\delta K(\overline{g},\,\delta g)$ vanish 
by definitions of these quantities (see Appendix~\ref{Appendix_Symplectic}),  
together with Eq.~(\ref{TT_cond}) for the timelike Killing field $\xi^M =(\p_t)^M$, and also by the fact that $\xi^M$ is vanishing at the bifurcate surface $\partial {\cal S}_\varepsilon$. 
Second, since the background spacetime is maximally symmetric, 
\begin{align}
E_\text{ct}^{\mu\nu}(\overline{g})\propto \overline{g}^{\mu\nu}, \qquad 
\tilde{E}_\text{bdy}^{\mu\nu}(\overline{\bdyg})\propto \overline{\bdyg}^{\mu\nu}
\end{align}
as follows from Eqs.~(\ref{Eij_ct}) and (\ref{Eij_bdy}). Therefore, their contractions with 
the traceless perturbations~(\ref{TT_cond}) vanish. 
According to the procedure in Ref.~\cite{Iyer:1994ys}, let us define the symplectic currents 
${\bm \omega}^{(d)}$, ${\bm \omega}_\text{ct}^{(d-1)}$, and ${\bm \omega}_\text{bdy}^{(d-1)}$ as 
\begin{eqnarray}
\label{symple_current} 
{\bm \omega}^{(d)}(\overline{G}, \delta G, {\cal L}_\xi G)
&=&
\delta {\bm \Theta}^{(d)}(\overline{G}, {\cal L}_\xi G)
-{\cal L}_\xi{\bm \Theta}^{(d)}(\overline{G},\delta G), 
 \nonumber \\
{\bm \omega}_\text{ct}^{(d-1)}(\overline{g}, \delta g, {\cal L}_\xi g)
 &=&
\delta {\bm \Theta}_\text{ct}^{(d-1)}(\overline{g}, {\cal L}_\xi g)
-{\cal L}_\xi{\bm \Theta}_\text{ct}^{(d-1)}(\overline{g},\delta g), 
\nonumber \\
{\bm \omega}_\text{bdy}^{(d-1)}(\overline{\bdyg}, \delta \bdyg, {\cal L}_\xi \bdyg)
 &=&\delta\tilde{\bm \Theta}_\text{bdy}^{(d-1)}(\overline{\bdyg}, {\cal L}_{\xi} \bdyg)
-{\cal L}_{\xi}\tilde{\bm \Theta}_\text{bdy}^{(d-1)}(\overline{\bdyg},\delta \bdyg). 
\end{eqnarray}
Since 
${\cal L}_\xi G_{MN}={\cal L}_\xi g_{\mu\nu}={\cal L}_\xi \bdyg_{\mu\nu}=0$, 
${\bm \omega}^{(d)}$, ${\bm \omega}^{(d-1)}_\text{ct}$, and ${\bm \omega}^{(d-1)}_\text{bdy}$ vanish. In addition, $\sigma^{(d-1)}$ and $\p{\cal S}^{(d-1)}_\varepsilon$ are the bifurcate 
Killing horizon with $\xi^M =0$. Therefore, by combining Eqs.~(\ref{J+j}) and (\ref{variation_J+j}), we obtain 
\begin{align}
\label{first_law}
\delta Q^\text{(ren)}_\sigma(\xi)+\delta\left(\int_{\p {\cal S}_\varepsilon}\tilde{\bm Q}_\text{bdy}^{(d-2)}(\tilde{\xi})\right)=0.  
\end{align}
The first term of the l.h.s. of eq.~(\ref{first_law}) corresponds to the variation of the holographic entanglement entropy. This follows from the Ryu-Takayanagi prescription~\cite{Ryu:2006bv}, since the bifurcation surface of the bulk Killing horizon is the corresponding extremal surface. The second term of the l.h.s. of eq.~(\ref{first_law}) corresponds to the variation of the gravitational (Wald) entropy of the horizon.

\section{Static perturbations of de Sitter spacetimes} 
We consider static perturbations satisfying the gauge condition as~(\ref{TT_cond}). Following the 
procedure of Ref.~\cite{Ishibashi:2025gvl}, we separate the perturbation $h_{\mu\nu}(x,z)$ in 
Eq.~(\ref{pertb_metric}) as $h_{\mu\nu}(x,z)=\xi(z)H_{\mu\nu}(x)$. Then, from the bulk 
Einstein equations, $E^{MN}_\text{bulk}=0$, derived from Eqs.~(\ref{variation_Lg}) reduce to
two decoupled equations: 
\begin{align}
\label{Eq_xi}
\xi''-\frac{(d-1)\Omega'}{\Omega}\xi'+m^2\xi=0, 
\end{align}
\begin{align}
\label{Eq_H}
\left(\overline{\bdyD}^2-\frac{2}{\ell^2}\right)H_{\mu\nu}(x)=m^2H_{\mu\nu}, 
\end{align}
where the mass-squared term $m^2$ appears as a separation constant. 
Hereafter, we focus on the case $m^2<0$, since, as shown below, only negative values of 
$m^2$ admit non-trivial static solutions.   

Imposing the regularity condition at $z=\infty$ for $\xi$ in (\ref{Eq_xi}), the solutions 
are expressed in terms of the hypergeometric function $F$ as  
\begin{subequations}
\label{xi_p}
\begin{align}
\label{xi_sol_dS}
& \xi=c(-y)^{\frac{d-1}{2}-p}F\left(p-\frac{d-1}{2},\,p+\frac{1}{2},\,2p+1;\frac{1}{y}   \right), \qquad 
y=\frac{1-\cosh \left(\frac{z}{\ell}\right)}{2}, 
\end{align}
\begin{align}
\label{def:p}
& p:=\sqrt{\frac{(d-1)^2}{4}-\hat{m}^2}, \qquad \hat{m}^2:=m^2\ell^2,  
\end{align}
\end{subequations}
where $c$ is an arbitrary constant. Under the above regularity condition at $y=-\infty$ (equivalently, $z=\infty$), 
a nontrivial bounded solution satisfying $|\xi|<\infty$ exists only for $m^2\le0$, whereas for $m^2>0$ the only 
regular solution is the trivial one, $\xi=0$.

We adopt the following metric ansatz for $H_{\mu\nu}$:
\begin{eqnarray}
\label{metric_ansatz_H}
 ds_d^2 &=&\bdyg_{\mu\nu}(u)dx^\mu dx^\nu \nonumber \\
&=&-\frac{f(u)}{u}(1+\epsilon T(u))dt^2+
\frac{\ell^2}{4u^2f(u)}(1+\epsilon U(u))du^2+\frac{\ell^2}{u}(1+\epsilon R(u))
d\Omega_{d-2}^2,   
\end{eqnarray}
where $1\le u < \infty$ and $u=\infty$ corresponds to the center of spherical symmetry. 

Under our gauge condition~(\ref{TT_cond}), i.e., $H^\mu{}_\mu=0=\bar{\cal D}^\nu H_{\mu \nu}$, Eqs.~(\ref{Eq_H}) reduce to the following set of equations:
\begin{subequations}
\label{H_equation}
\begin{align}
\label{Eq_R}
R=-\frac{T+U}{d-2}, 
\end{align}
\begin{align}
\label{Eq_T}
T=\left(\frac{d}{u}-(d-1) \right)U+2(u-1)U', 
\end{align}
\begin{align}
\label{Eq_U}
U''-\left(\frac{d+1}{2u}+\frac{2}{1-u}  \right)U'+\frac{2d+2+\hat{m}^2}{4u^2(1-u)}U=0. 
\end{align}
\end{subequations}
Imposing the regularity at $u=\infty$, the solution of Eq.~(\ref{Eq_U}) is given by the hypergeometric 
function 
\begin{align}
\label{sol_U}
U=F\left(\frac{d+3}{4}+\frac{p}{2}, \frac{d+3}{4}-\frac{p}{2}, \frac{d+1}{2}, \frac{1}{u}  \right) .
\end{align}
Making use of the hypergeometric identity 
\begin{eqnarray}
 F(\alpha,\beta,\gamma;z)
 &=&\frac{\Gamma(\alpha+\beta-\gamma)\Gamma(\gamma)}
{\Gamma(\alpha)\Gamma(\beta)}(1-z)^{\gamma-\alpha-\beta}
F(\gamma-\alpha,\gamma-\beta, \gamma-\alpha-\beta +1;1-z) \nonumber \\
&+&\frac{\Gamma(\gamma-\alpha-\beta)\Gamma(\gamma)}
{\Gamma(\gamma-\alpha)\Gamma(\gamma-\beta)}
F(\alpha, \beta, \alpha+\beta-\gamma+1;1-z), 
\end{eqnarray}
the regularity at the de Sitter horizon~($u=1$) requires 
\begin{align}
\label{pn_relation}
\frac{d+3}{4}-\frac{p}{2}=-n, \qquad n=0,1,2,\cdots. 
\end{align}
 Eq.~(\ref{def:p}) implies that the largest value of $\hat{m}^2$ admitting a regular static solution is 
attained for $n=0$, namely
\begin{align}
\label{L_mass}
\hat{m}^2=-2(d+1), 
\end{align}
which agrees with the threshold mass for the instability of de Sitter spacetime obtained in
Ref.~\cite{Ishibashi:2025gvl}. 
 
Substituting $n=0$ into Eqs.~(\ref{sol_U}), (\ref{Eq_R}), (\ref{Eq_T}) yields  
\begin{align}
\label{sol_H}
U=1, \qquad T=\frac{d}{u}+1-d, \qquad R=1-\frac{d}{(d-2)u}, 
\end{align} 
where the overall normalization has been fixed so that $U=1$. 

For later convenience, it is useful to express the bulk Noether charge $\int {\bm Q}^{(d-1)}$ with 
the $(d-1)$-form, (\ref{bulk_NC_form}), in terms of the functions $U$, $T$, and $R$. For each 
surface $\sigma^{(d-1)}$ and ${\cal S}^{(d-1)}_\varepsilon$, they are written as 
\begin{eqnarray}
\label{Bulk_NC_H}
Q_\sigma &:=&\int_\sigma {\bm Q}^{(d-1)}
=\frac{\ell^{d-3}}{8\pi G_{d+1}}\int_\sigma \frac{1-\epsilon \xi(z)}{\Omega^{d-1}(z)}dz
d\Omega_{d-2}
\nonumber \\
&=&\frac{\ell^{d-3}}{4 G_{d+1}}\frac{\pi^\frac{d-3}{2}}{\Gamma\left(\frac{d-1}{2} \right)}
\int_\varepsilon^\infty \frac{1-\epsilon \xi(z)}{\Omega^{d-1}(z)}dz, 
\\ 
\label{Bulk_NC_E}
Q_\text{E}&:=&\int_{{\cal S}_\varepsilon} {\bm Q}^{(d-1)} 
 =\int_{{\cal S}_\varepsilon}  m^N n^M Q_{MN}
 \nonumber \\
&=&\frac{\ell^{d-1}}{16 G_{d+1}}\frac{\pi^\frac{d-3}{2}}{\Gamma\left(\frac{d-1}{2} \right)}
\int^\infty_1 \frac{\Omega^{-d}}{u^\frac{d+1}{2}}
(-2 \partial_z \Omega +\epsilon\Omega (\partial_z \xi) T)du\Biggr{|}_{z=\varepsilon}, 
\end{eqnarray}
where $\Gamma(z)$ denotes the Gamma function and where in the second integral 
of (\ref{Bulk_NC_E}), the volume element of ${\cal S}_\varepsilon^{(d-1)}$ is understood. 
Here $n^M$ and $m^M$ are, respectively, the unit (future directed) timelike and spacelike (outward) vectors orthogonal to ${\cal S}^{(d-1)}_\varepsilon$. 
It is also useful to evaluate the second term in Eq.~(\ref{def:ren_charge_H}) under the metric ansatz~(\ref{metric_ansatz_H}). Let us define a unit spacelike vector $s^\mu$ orthogonal to 
$\p {\cal S}_\varepsilon$ as 
\begin{align}
\label{vec_s}
 s^M =-\frac{2u\sqrt{f}\Omega}{\ell}(1+\epsilon \xi U)^{-\frac{1}{2}} (\p_u)^M. 
\end{align}
With this definition, we find $\int_{\partial {\cal S}_\varepsilon} {\bm Q}_\text{ct}^{(d-2)}$ to be written as 
\begin{eqnarray}
\label{NC_ct}
 Q_\text{ct}
 &:=& \int_{\p {\cal S}_\varepsilon} s_\nu n_\mu Q^{\mu\nu}_\text{ct}
 \nonumber \\ 
 &=&
-\frac{L\ell^{d-3}}{8\pi(d-2) G_{d+1}\Omega^{d-2}(\varepsilon)}\int
(1-\epsilon\xi(\varepsilon))d\Omega_{d-2} 
\nonumber \\
&{}& -\frac{c_dL^3\ell^{d-3}}{8\pi(d-2)^2 G_{d+1}\Omega^{d-2}(\varepsilon)}\int
(1-\epsilon\xi(\varepsilon))\left(2{R^u}_u-\frac{d}{2(d-1)}R\right)d\Omega_{d-2}\Biggr{|}_{u=1}. 
\end{eqnarray} 
where in the last equality, 
we used $f(1)=0$, $U(1)=T(1)=1$, and ${H^\mu}_\mu=0$ as follows from Eqs.~(\ref{sol_H}). 
In the remainder of this paper, we take the limit $\varepsilon\to 0$ for
the renormalized charges~(\ref{def:ren_charge_H}) and (\ref{def:ren_charge_E}).   

\subsection{$d=3$ de Sitter spacetime}
Let us first consider the simplest case, $d=3$, in which the higher curvature corrections are absent. 
In this case, substituting Eq.~(\ref{L_mass}) into Eq.~(\ref{xi_sol_dS}), we obtain 
\begin{align}
\label{sol_xi_3D}
\xi=2e^{-\frac{2z}{\ell}}-e^{-\frac{4z}{\ell}}. 
\end{align}  
By Eqs.~(\ref{sol_H}), (\ref{Ricci_bulk}), and (\ref{sol_xi_3D}), the variations of the charges in 
Eqs.~(\ref{Bulk_NC_H}), (\ref{Bulk_NC_E}), and (\ref{NC_ct}) are 
\begin{align}
\label{QdS3_h}
& \delta Q_\sigma=\frac{3\epsilon L^2}{4G_4\ell}-\frac{\epsilon L^2}{4G_4\varepsilon}+O(\varepsilon), 
\nonumber \\
&\delta Q_\text{ct}=\frac{\epsilon L^2}{4G_4\varepsilon}+O(\varepsilon),  \nonumber \\
& \delta Q_\text{E}=\frac{\epsilon \ell^2}{16 G_4}
\int^\infty_1 \frac{\Omega^{-2}(\varepsilon)}{u^2}(\partial_z\xi) T(u) du
=-\frac{3\epsilon L^2}{4G_4\ell}+\frac{\epsilon L^2}{4G_4\varepsilon}+O(\varepsilon). 
\end{align}   
By Eq.~(\ref{TT_cond}), the variation of the second term in Eq.~(\ref{def:ren_charge_E}) vanishes and 
the variation of the third term in Eq.~(\ref{def:ren_charge_E}) yields 
\begin{align}
-2\delta\int_{{\cal S}_\varepsilon} n_\mu {{E_\text{ct}}^\mu}_\nu\xi^\nu=
-\frac{L}{8\pi G_4}\int_{{\cal S}_\varepsilon}
{\bm \epsilon}_{\mu\alpha\beta}\delta {R^\mu}_\nu \xi^\nu=-\frac{\epsilon L^2}{4G_4\varepsilon}
+O(\varepsilon), 
\end{align}
where we used Eqs.~(\ref{L_mass}), (\ref{sol_H}), and (\ref{Ricci_bulk}). Thus, the renormalized charges 
in Eqs.~(\ref{def:ren_charge_H}) and (\ref{def:ren_charge_E}) are finite, and we obtain
\begin{align}
\label{devi_3D_HE}
\delta Q^{(\text{ren})}_\sigma =
-\delta Q^{(\text{ren})}_\text{E}=\frac{3\epsilon L^2}{4G_4\ell}. 
\end{align}
 
The variation of the gravitational charge on the boundary horizon $u=1$ is evaluated 
by Eqs.~(\ref{anti_symmet_tensor_bdy}) and (\ref{sol_H}), with $\alpha_1=\alpha_2=\alpha_3=0$, as 
\begin{align}
\label{grav_charge_3}
\delta \int_{\p {\cal S}_\varepsilon}\tilde{\bm Q}_\text{bdy}^{(d-2)}
=\delta\int_{\p {\cal S}_\varepsilon} \tilde{s}_\nu \tilde{n}_\mu Q^{\mu\nu}_\text{bdy}=-\frac{\epsilon}{4G_3},  
\end{align}
where $\tilde{s}^\mu$ and $\tilde{n}^\mu$ are unit spacelike and timelike vectors orthogonal to $\p {\cal S}_\varepsilon$ 
associated with the conformal metric~(\ref{metric_ansatz_H}), 
\begin{align}
\label{vec_s}
\tilde{s}^\mu=-\frac{2u\sqrt{f}}{\ell}(1+\epsilon U)^{-{1}/{2}} (\p_u)^\mu, \qquad 
\tilde{n}^\mu=\sqrt{\frac{u}{f}}(1+\epsilon T)^{-{1}/{2}}(\p_t)^\mu.  
\end{align}
The previous work~\cite{Ishibashi:2025gvl} showed that 
the SCE Eqs.~(\ref{semi_Eqs}) are satisfied if the following algebraic relation holds for $\hat{m}^2=-8$~(\ref{L_mass}):
\begin{align}
\label{algebraic_3}
\gamma_3:=\frac{L^2G_3}{\pi G_4\ell}=\frac{1}{\pi\sqrt{1-\hat{m}^2}}=\frac{1}{3\pi}. 
\end{align}
Thus, the variation of the gravitational charge~(\ref{grav_charge_3}) is given by
\begin{align}
\label{H_E_relation3}
\delta \int_{\p {\cal S}_\varepsilon}\tilde{\bm Q}_\text{bdy}^{(d-2)} 
 =-\frac{\epsilon}{4G_3}=-\delta Q^{(\text{ren})}_\sigma 
\end{align}
together with Eq.~(\ref{devi_3D_HE}), confirming (\ref{first_law}) for the $d=3$ case. 

\subsection{$d=5$ de Sitter spacetime}
Next, we consider the $d=5$ de Sitter case. Substituting $p=4$, which follows from Eq.~(\ref{L_mass}), into 
Eq.~(\ref{xi_sol_dS}), we obtain 
\begin{align}
\label{sol_xi_5D}
\xi(z)=\frac{5}{32}(-y(z))^{-2}F\left(2,\,\frac{9}{2},\, 9, \,\frac{1}{y(z)}\right)=\frac{1}{2}e^{-\frac{6z}{\ell}}
\left(1-4e^{\frac{2z}{\ell}}+5e^{\frac{4z}{\ell}}\right), 
\end{align}
where $\xi$ is normalized so that $\xi(0)=1$. 
By Eqs.~(\ref{sol_H}), (\ref{Ricci_bulk}), and (\ref{sol_xi_5D}), the variations of the charges defined in 
Eqs.~(\ref{Bulk_NC_H}), (\ref{Bulk_NC_E}), and (\ref{NC_ct}) are 
\begin{align}
\label{QdS5_h}
& \delta Q_\sigma =-\frac{\epsilon \pi \ell^2}{4G_6}\int^\infty_\varepsilon \frac{\xi(z)}{\Omega^4(z)}dz
=-\frac{\pi L^4\ell^2\epsilon}{12G_6\varepsilon^3}+\frac{2\pi L^4\epsilon}{3G_6\varepsilon}
-\frac{5\pi L^4\epsilon}{3G_6\ell}
+O(\varepsilon), \nonumber \\
& 
\delta Q_\text{ct}=\frac{\pi\epsilon L^4\xi(\varepsilon)}{12G_6 \ell\sinh^3({\varepsilon}/{\ell})}
 - \frac{11\pi\epsilon L^4\xi(\varepsilon)}{24G_6 \ell\sinh({\varepsilon}/{\ell})}=
\frac{\pi L^4\ell^2 \epsilon}{12G_6\varepsilon^3}-\frac{2\pi L^4\epsilon}{3G_6\varepsilon}+O(\varepsilon),  \nonumber \\
& 
\delta Q_\text{E}=\frac{\epsilon \pi\ell^4}{16G_6}
\int^\infty_1 \frac{\Omega^{-4}(\varepsilon)}{u^3}
 \partial_z \xi(\varepsilon)T(u)du=\frac{\pi L^4\ell^2\epsilon}{12G_6\varepsilon^3}-\frac{2\pi L^4\epsilon}{3G_6\varepsilon}
+\frac{5\pi L^4\epsilon}{3G_6\ell}+O(\varepsilon). 
\end{align}   
By Eq.~(\ref{TT_cond}), the variation of the second term in Eq.~(\ref{def:ren_charge_E}) vanishes, while 
the variation of the third term in Eq.~(\ref{def:ren_charge_E}) yields 
\begin{eqnarray}
-2\delta\int_{{\cal S}_\varepsilon} n_\mu {{E_\text{ct}}^\mu}_\nu \xi^\nu
 &=& -\frac{\pi \epsilon L^4\hat{m}^2}{24G_6 \ell\,\sinh^3({\varepsilon}/{\ell})}\int^\infty_1
\left[\frac{1}{2}+\left(\frac{\hat{m}^2}{6}-\frac{3}{4}\right)\sinh^2({\varepsilon}/{\ell}) \right]\, 
\frac{\xi(\varepsilon) T(u)}{u^3} du \nonumber \\
 &=& -\frac{\pi L^4\ell^2\epsilon}{12G_6\varepsilon^3}+\frac{2\pi L^4\epsilon}{3G_6\varepsilon}+O(\varepsilon),  
\end{eqnarray}
where we used Eqs.~(\ref{L_mass}), (\ref{sol_H}), (\ref{Ricci_bulk}), (\ref{H1_pert_dS}), and (\ref{H2_pert_dS}). 
Thus, the renormalized charges in Eqs.~(\ref{def:ren_charge_H}) and (\ref{def:ren_charge_E}) are finite, and we obtain
\begin{align}
\label{H_E_relation5}
\delta Q^{(\text{ren})}_\sigma =
-\delta Q^{(\text{ren})}_\text{E}=-\frac{5\pi L^4\epsilon}{3G_6\ell}. 
\end{align}

The variation of the gravitational charge on the boundary horizon $u=1$ is evaluated 
using Eqs.~(\ref{anti_symmet_tensor_bdy}) and (\ref{sol_H}) as 
\begin{align}
\label{grav_charge_5}
\delta \int_{\p {\cal S}_\varepsilon}\tilde{\bm Q}_\text{bdy}^{(d-2)}
=\delta\int_{\p {\cal S}_\varepsilon} \tilde{s}_\nu \tilde{n}_\mu Q^{\mu\nu}_\text{bdy}
=-\frac{\pi\epsilon \ell^2}{4G_5}(1+40\hat{\alpha}_1-4\hat{\alpha}_2-52\hat{\alpha}_3) , 
\end{align}
where $\tilde{s}^\mu$ and $\tilde{n}^\mu$ are defined in Eq.~(\ref{vec_s}).
As shown in Ref.~\cite{Ishibashi:2025gvl}, 
the following algebraic relation follows from the SCE Eqs.~(\ref{semi_Eqs}):
\begin{align}
\label{algebraic_5}
\gamma_5:=\frac{G_5L^4}{\pi\ell^3G_6}
=-\frac{3(1+40\hat{\alpha}_1-4\hat{\alpha}_2-52\hat{\alpha}_3)}{20\pi}, 
\end{align}
where we have used $\hat{m}^2=-12$~(\ref{L_mass}) in the last 
equality~\footnote{Here, $\gamma_5>0$. However, note that the validity of Eq.~(\ref{first_law}) 
does not depend on the sign of $\gamma_5$.}.
Combining Eqs.~(\ref{H_E_relation5}), (\ref{grav_charge_5}), and (\ref{algebraic_5}), we 
confirm Eq.~(\ref{first_law}) for the $d=5$ case.

\subsection{$d=4$ de Sitter spacetime}
In the $d=4$ case, the calculations of the renormalized charges in Eqs.~(\ref{def:ren_charge_H}) and (\ref{def:ren_charge_E}) are somewhat more involved due to the presence of the Weyl anomaly.

Substituting $p=7/2$, which follows from Eq.~(\ref{L_mass}), into Eq.~(\ref{xi_sol_dS}), we obtain 
\begin{align}
\label{sol_xi_4D}
\xi(y)=1-10y(y-1)\{1+12y(y-1) \}-60y^2(y-1)^2(2y-1)\ln\left(1-\frac{1}{y}\right), 
\end{align}
where $y$ is a function of $z$, as defined by Eq.~(\ref{xi_sol_dS}), and  
$\xi$ is normalized so that $\xi(0)=1$. Substituting (\ref{sol_xi_4D}) into the expression for the variation of the charge (\ref{Bulk_NC_H}), we obtain 
\begin{eqnarray}
\label{variation_Q_H4} 
\delta Q_\sigma 
 &=& -\frac{\epsilon L^3}{2G_5}\int^\infty_\varepsilon \frac{\xi}{\ell^2\sinh^3(z/\ell)}dz
=-\frac{\epsilon L^3}{16\ell G_5}\int^{-\delta_\varepsilon}_{-\infty}\frac{\xi(y)dy}{y^2(y-1)^2} \nonumber \\
&=&-\frac{\epsilon L^3}{16\ell G_5}
\left[30+\frac{1}{\delta_\varepsilon}+60\delta_\varepsilon+\frac{1}{1+\delta_\varepsilon}
-12(1+5\delta_\varepsilon+5\delta_\varepsilon^2)
\ln \left(1+\frac{1}{\delta_\varepsilon}\right)\right] \nonumber \\
&=&-\frac{\epsilon L^3}{2G_5\ell}
\left[\frac{23}{6}+\frac{\ell^2}{2\varepsilon^2}+3\ln \left(\frac{\varepsilon}{2\ell}\right)\right]+O(\varepsilon), 
\end{eqnarray}
where $\delta_\varepsilon=-y(\varepsilon)>0$, and we have used 
$\delta_\varepsilon\simeq \varepsilon^2/(4\ell^2)+\varepsilon^4/(48\ell^4)$ in the last equality. 
Using Eq.~(\ref{Ricci_bulk}), the variation of the charge in Eq.~(\ref{NC_ct}) is given by 
\begin{align}
\label{variation_Q_ct4}
\delta Q_\text{ct}=\frac{\epsilon L^3\xi(\varepsilon)}{4G_5 \ell}
\left(\frac{1}{\sinh^2({\varepsilon}/{\ell})}+6\ln(\mu \varepsilon)\right). 
\end{align} 
Adding Eqs.~(\ref{variation_Q_H4}) and (\ref{variation_Q_ct4}), the divergences cancel, yielding the finite variation 
of the renormalized charge $Q_\sigma$~(\ref{def:ren_charge_H}): 
\begin{eqnarray}
\label{vari_renQ_H4}
\delta Q^{(\text{ren})}_\sigma 
 &=&
\lim_{\varepsilon\to 0}\frac{\epsilon L^3}{4G_5\ell}\left[-
\left\{\frac{23}{3}+\frac{\ell^2}{\varepsilon^2}+6\ln \left(\frac{\varepsilon}{2\ell}\right)\right\}
+\left(\frac{1}{\sinh^2({\varepsilon}/{\ell})}+6\ln(\mu \varepsilon)\right) \right] \nonumber \\
 &=&\frac{3\epsilon L^3}{2G_5 \ell}\left(\ln(2\mu \ell)-\frac{7}{4}   \right). 
\end{eqnarray}
As before, by Eq.~(\ref{TT_cond}), the variation of the second term in Eq.~(\ref{def:ren_charge_E}) vanishes. The remaining terms can be evaluated using Eqs.~(\ref{L_mass}), (\ref{sol_H}), (\ref{sol_xi_4D}), and (\ref{H_pert_dS}), giving  
\begin{align}
\label{vari_renQ_E4}
&\delta Q^{(\text{ren})}_\text{E}=
\lim_{\varepsilon\to 0}\int^\infty_1 \frac{T}{u^\frac{5}{2}}du\times \nonumber \\
& \qquad \qquad \frac{\epsilon \ell^3}{8G_5}
\left[\Omega^{-3}(\varepsilon)\xi_{,z}(\varepsilon)-2\xi(\varepsilon)
\left(\frac{Lm^2}{4\Omega^2(\varepsilon)}-L^3\ln(\mu \varepsilon)
\left\{\frac{m^2}{8}\left(m^2+\frac{6}{\ell^2}\right)-\frac{m^2}{\ell^2}\right\}\right)   \right]
\nonumber \\
&\qquad =\frac{3\epsilon L^3}{8G_5\ell}(7-4\ln(2\mu \ell)). 
\end{align}
Thus, we have confirmed that the variations of both renormalized charges, $\delta Q^{(\text{ren})}_{\sigma}$ and $\delta Q^{(\text{ren})}_\text{E}$, are finite. 

The variation of the gravitational charge on the boundary horizon $u=1$ is evaluated 
using Eqs.~(\ref{anti_symmet_tensor_bdy}) and (\ref{sol_H}), yielding  
\begin{align}
\label{grav_charge_4}
\delta \int_{\p {\cal S}_\varepsilon}\tilde{\bm Q}_\text{bdy}^{(d-2)}
=-\frac{\epsilon \ell}{2G_4}\left(1+24\hat{\alpha}_1-4\hat{\alpha}_2-40\hat{\alpha}_3   \right), 
\end{align}
where $\tilde{s}^\mu$ and $\tilde{n}^\mu$ are defined in Eq.~(\ref{vec_s}).
As shown in Ref.~\cite{Ishibashi:2025gvl}, the first-order variations of $\pExp{{{\cal T}_\mu}^\nu}$ and 
${{\cal E}_\mu}^\nu$ in the SCE Eqs.~(\ref{semi_Eqs}) are 
\begin{eqnarray}
\label{4dim_SE}
 \delta \pExp{{{\cal T}_\mu}^\nu} &=&-\frac{\epsilon L^3\hat{m}^2}{64\pi G_5\ell^4}
\left[1+(\hat{m}^2-2)\left\{\psi\left(p+\frac{1}{2}\right)+\gamma_E-\ln(2\mu \ell)   \right\}    \right]
{H_\mu}^\nu 
\nonumber \\
  &=& \frac{5\epsilon L^3}{32\pi G_5\ell^4}(-21+12\ln(2\mu \ell)){H_\mu}^\nu, 
\end{eqnarray}
\begin{align}
\label{4dim_curvature}
\delta ({\cal E}_{\mu\rho}\bdyg^{\rho\nu})=\frac{5\epsilon}{\ell^2}
\left(1+24\hat{\alpha}_1-4\hat{\alpha}_2-40\hat{\alpha}_3   \right){H_\mu}^\nu, 
\end{align}
where $\psi(x)$ and $\gamma_E$ are the digamma function and Euler constant, and 
we used 
\begin{align}
\psi\left(4\right)+\gamma_E=\frac{11}{6}
\end{align}
corresponding to $p=7/2$. Substituting Eqs.~(\ref{4dim_SE}) and (\ref{4dim_curvature}) into Eqs.~(\ref{semi_Eqs}), 
we obtain the following algebraic equation
\begin{align}
\label{algebra_4dim}
4\left(1+24\hat{\alpha}_1-4\hat{\alpha}_2-40\hat{\alpha}_3   \right)=
\frac{G_4L^3}{G_5\ell^2}(-21+12\ln(2\mu \ell)). 
\end{align}
Using Eq.~(\ref{algebra_4dim}) to eliminate $G_4$ from Eq.~(\ref{grav_charge_4}), we find that the variation of the 
gravitational charge coincides with the variation of the renormalized charge $\delta Q^{(\text{ren})}_{\rm E}$. Together with Eqs.~(\ref{vari_renQ_H4}) and (\ref{vari_renQ_E4}), this confirms Eq.~(\ref{first_law}) for the $d=4$ case.

\section{Summary and discussions}
We have studied the generalized de Sitter entropy, which consists of the entanglement entropy of the strongly coupled quantum field theory and the gravitational Wald entropy, in the framework of holographic semiclassical gravity~\cite{Compere:2008us, Ishibashi:2023luz}. For static perturbations, we have shown that the first order variation of the entanglement entropy is finite without introducing a cutoff, by applying the Noether charge formula to the semiclassical model, in which the perturbation does not decay as rapidly as required by the Dirichlet boundary condition. This finiteness arises from the cancellation between the divergent contributions to the area of the bulk horizon and the Noether charges associated with the boundary counterterm action. Using the semiclassical Einstein equations~(\ref{semi_Eqs}), we have further shown that the sum of the first-order variations of the entanglement entropy and the gravitational Wald entropy vanishes. This result is naturally understood in terms of the first 
laws for the gravitational entropy of the boundary de Sitter horizon and the entanglement entropy of the bulk region associated with that horizon, as explained in the Introduction. 
   
It would be interesting to extend our analysis to other semiclassical spacetimes, including those containing black holes. For example, in the case of an AdS black hole, which provides an example of a boundary semiclassical spacetime, one can identify three quantities associated with Noether charges: the canonical energy evaluated at the asymptotic boundary of the boundary spacetime, the bulk entanglement entropy, and the gravitational entropy of the black hole horizon. Although the boundary field decays more slowly than in the Dirichlet case due to the presence of a negative mass term, we expect that the canonical energy can still be rendered finite by introducing an appropriate counterterm~\cite{Ishibashi:2024fnm}. In this case, we expect to obtain a first law in which the variation of the total energy of the boundary spacetime is related to the variation of the generalized entropy. We leave a detailed investigation of this possibility to future work.   
     
We expect the first laws associated with the gravitational entropy of the boundary de Sitter horizon and the entanglement entropy of the corresponding bulk region to hold even for dynamical perturbations. This suggests that the first-order variation of the generalized entropy vanishes, corresponding to a saturation of the generalized second law at first order.
From the bulk perspective, the entanglement entropy is related to the area of the bulk horizon. By introducing a timelike cutoff hypersurface near the AdS boundary~\cite{Bunting:2015sfa}, the change in the bulk horizon area along the horizon null generators can be understood in terms of the flux of the null generators across this cutoff hypersurface.  In this picture, the area variation can be decomposed into a flux contribution and a nonnegative contribution associated with the classical area increase of the bulk horizon. The former arises from the flux of the horizon generators across the cutoff hypersurface, while the latter reflects the area increase governed by the classical area theorem for black holes~\cite{HawkingEllis}. 
For perturbations about a stationary horizon, the terms quadratic in the expansion and shear in the Raychaudhuri equation contribute at second order in perturbation theory. These terms govern the nonlinear contribution to the horizon area change and underlie the classical area increase theorem, subject to the appropriate assumptions.

In our semiclassical model, the bulk entanglement entropy is expected to remain finite without introducing a cutoff hypersurface. It would be interesting to investigate whether the renormalized generalized entropy obeys a second law in our semiclassical model, despite the possible increase or decrease of the bulk horizon area due to the flux of the horizon generators across the cutoff hypersurface.      
     
\begin{center}
{\bf Acknowledgments}
\end{center} 
We wish to thank Gary T. Horowitz and Donald Marolf for useful discussions. 
This work is supported in part by JSPS KAKENHI Grant No. 25K07306 (K.M.) and also in part supported by MEXT KAKENHI Grant-in-Aid for Transformative Research Areas
A Extreme Universe No.21H05186 (A.I. and K.M.) and 21H05182 and JSPS KAKENHI Grant No. JP26K07105. 

\appendix
\section{Symplectic potential and the Noether current}
\label{Appendix_Symplectic}
Here we collect several formulas. The variations are taken for the lower-index quantities. However, for simplicity of notation, we often use mixed indices. For instance, $\delta G^M{}_N$ implies $g^{ML} \delta G_{LN}$ rather than $\delta (g^{ML} G_{LN})$.
${\bm \Theta}^{(d)}$, ${\bm \Theta}_\text{ct}^{(d-1)}$, and ${\bm \Theta}_\text{bdy}^{(d-1)}$ are derived from 
the action~(\ref{bulk_action}) and (\ref{bdy_action}) as 
\begin{subequations}
\label{symplectic_form}
\begin{eqnarray}
\label{symplec_bulk}
{{\bm \Theta}}_{K_2\cdots K_{d+1}}=\dfrac{{\bm \epsilon}_{MK_2\cdots K_{d+1}}}{16\pi G_{d+1}}
\left(\nabla^N \delta {G^M}_N-\nabla^M {\delta G^N}_N  \right), 
\end{eqnarray}
\begin{eqnarray}
\label{symplec_bdy}
 \tilde{\bm \Theta}^\text{bdy}{}_{\nu_2\cdots \nu_d}=
 \tilde{\bm\epsilon}_{\mu\nu_2\cdots \nu_d}
 \left(\Theta^\mu_{(0)}+\sum_{i=1}^3\alpha_i \Theta^\mu_{(i)}  \right),   
\end{eqnarray}
\end{subequations}
where $\Theta^\mu_{(0)}$ and $\Theta^\mu_{(i)}$ are defined by 
\begin{subequations}
\label{bdy_sympl_form}
\begin{eqnarray}
\label{bdy_sympl_form0}
\Theta^\mu_{(0)}=\dfrac{1}{16\pi G_d}(\bdyD^\nu \delta \bdyg^\mu_\nu-\bdyD^\mu \delta \bdyg), 
\end{eqnarray}
\begin{eqnarray}
\label{bdy_sympl_form1}
\Theta_{(1)}^\mu=\dfrac{1}{16\pi G_d}\left[-2(\bdyD^\nu \bdyR)\delta {\bdyg^\mu}_\nu+2(\bdyD^\mu\bdyR)\delta\bdyg
+2\bdyR \bdyD^\nu \delta {\bdyg^\mu}_\nu-2\bdyR \bdyD^\mu \delta \bdyg\right], 
\end{eqnarray}
\begin{eqnarray}
\label{bdy_sympl_form2}
&\Theta_{(2)}^\mu=
\dfrac{1}{16\pi G_d}
\Biggl[-\bdyR^{\rho\nu}\bdyD^\mu{\delta \bdyg}_{\rho\nu}
+\delta \bdyg_{\rho\nu}\bdyD^\mu\bdyR^{\rho\nu}
-\bdyR^{\mu\rho} \bdyD_\rho \delta \bdyg
+(\bdyD_\rho \bdyR^{\rho\mu})\delta \bdyg \nonumber \\
&+2\bdyR^{\rho\nu} \bdyD_\rho {{\delta \bdyg}^\mu}_\nu
-2{{\delta \bdyg}^\rho}_\nu\bdyD_\rho \bdyR^{\mu\nu}\Biggr], 
\end{eqnarray}
\begin{eqnarray}
\label{bdy_sympl_form3}
\Theta_{(3)}^\mu=\dfrac{1}{4\pi G_d}\left({\bdyR^\mu}_{\nu\alpha\beta}\bdyD^\beta\delta \bdyg^{\alpha\nu}
-\delta\bdyg^{\alpha\nu}\bdyD^\rho{\bdyR_{\rho\nu\alpha}}^\mu\right). 
\end{eqnarray}
\end{subequations}

For an arbitrary vector field $\xi^\mu$, substituting 
\begin{eqnarray}
\label{Lie_derivative}
\delta \bdyg_{\mu\nu}=\bdyD_\mu \xi_\nu+\bdyD_\nu \xi_\mu
\end{eqnarray}
into (\ref{bdy_sympl_form}), one obtains 
\begin{subequations}
\label{symplec_Lie}
\begin{eqnarray}
\label{symplec_Lie0}
\Theta_{(0)}^\mu
=\dfrac{\bdyD_\nu(\bdyD^\nu \xi^\mu-\bdyD^\mu \xi^\nu)}{16\pi G_d}+\frac{{\bdyR^\mu}_\alpha\xi^\alpha}{8\pi G_d}, 
\end{eqnarray}
\begin{eqnarray}
\label{symplec_Lie1}
& \Theta_{(1)}^\mu=\dfrac{1}{4\pi G_d}\left[\bdyD^2\bdyR\,{\delta^\mu}_\nu-\bdyD^\mu\bdyD_\nu\bdyR
+\bdyR{\bdyR^\mu}_\nu\right]\xi^\nu 
+\dfrac{1}{4\pi G_d}D_\nu\left(\bdyR\bdyD^{[\nu} \xi^{\mu]}-2\bdyD^{[\nu}\bdyR\,\xi^{\mu]}\right), 
\end{eqnarray}
\begin{eqnarray}
\label{symplec_Lie2}
&\Theta_{(2)}^\mu=
\dfrac{1}{16\pi G_d}[2\bdyD^2 {\bdyR^\mu}_\nu-2\bdyD^\mu\bdyD_\nu \bdyR+\bdyD^2\bdyR\, {\delta^\mu}_\nu
+4{\bdyR^\mu}_{\rho\nu\beta}\bdyR^{\rho\beta}]\xi^\nu 
\nonumber \\
& \qquad \quad +\dfrac{1}{8\pi G_d}\bdyD_\nu\left(\bdyD^{[\rho}\xi^{\mu]}{\bdyR^\nu}_\rho
-\bdyD^{[\rho}\xi^{\nu]}{\bdyR^\mu}_\rho-\xi^{[\mu}\bdyD^{\nu]}\bdyR+2\xi^\rho\bdyD^{[\mu}{\bdyR^{\nu]}}_\rho \right), 
\end{eqnarray}
\begin{eqnarray}
\label{symplec_Lie3}
& \Theta_{(3)}^\mu=\dfrac{1}{8\pi G_d}\Biggl[-2\bdyD^\mu\bdyD_\nu\bdyR+4{\bdyR^\mu}_{\rho\nu\beta}\bdyR^{\rho\beta}
-4{\bdyR_\nu}^\rho{\bdyR^\mu}_\rho+4\bdyD^2{\bdyR^\mu}_\nu+
2{\bdyR^\mu}_{\gamma\beta\rho}{\bdyR_\nu}^{\gamma\beta\rho}\Biggr]\xi^\nu
\nonumber \\ 
&+\dfrac{1}{4\pi G_d}\bdyD_\nu\Bigl({\bdyR_{\beta\rho}}^{[\nu\mu]}\bdyD^\beta\xi^\rho
+4\xi^\rho\bdyD^{[\mu}{\bdyR^{\nu]}}_\rho  \Bigr).  
\end{eqnarray}
\end{subequations}
This yields
\begin{eqnarray}
\label{Theta_bdy}
\tilde{\bm \Theta}^\text{bdy}{}_{\nu_2\cdots \nu_d}=
\tilde{\bm\epsilon}_{\mu\nu_2\cdots \nu_d}
\left(\bdyD_\nu Q^{\mu\nu}_\text{bdy}+\frac{{{\cal E}^\mu}_\nu\xi^\nu}{8\pi G_d} 
 + L_\text{bdy}\xi^\mu \right)
\end{eqnarray}
where $Q^{\mu\nu}_\text{bdy}$ is defined by 
\begin{eqnarray}
\label{anti_symmet_tensor_bdy}
 Q^{\mu\nu}_\text{bdy}
 &:=&-\dfrac{1}{8\pi G_d}
\Bigl[  
\bdyD^{[\mu} \xi^{\nu]}-2\alpha_1(\bdyR\bdyD^{[\nu} \xi^{\mu]}-2\bdyD^{[\nu}\bdyR\,\xi^{\mu]}) 
\nonumber \\
&{}& -\alpha_2\left(\bdyD^{[\rho}\xi^{\mu]}{\bdyR^\nu}_\rho
-\bdyD^{[\rho}\xi^{\nu]}{\bdyR^\mu}_\rho-\xi^{[\mu}\bdyD^{\nu]}\bdyR+2\xi^\rho\bdyD^{[\mu}{\bdyR^{\nu]}}_\rho\right)  \nonumber \\
&{}& +2\alpha_3\left( {\bdyR_{\beta\rho}}^{[\mu\nu]}\bdyD^\beta\xi^\rho
-4\xi^\rho\bdyD^{[\mu}{\bdyR^{\nu]}}_\rho   \right)  \Bigr]. 
\end{eqnarray}

Similarly, we first have 
\begin{eqnarray}
\label{Theta_ct}
{{\bm \Theta}^\text{ct}}_{\nu_2\cdots \nu_d}={\bm\epsilon}_{\mu\nu_2\cdots \nu_d}
\left\{D_\nu Q^{\mu\nu}_\text{ct}-2E_\text{ct}{}^\mu{}_\nu \xi^\nu+
 L_\text{ct}\xi^\mu\right\}. 
\end{eqnarray}
Then, $Q^{\mu\nu}_\text{ct}$ and the rest of the corresponding formulas for the counter-term can be obtained from the above quantities by the following replacements, 
\begin{align*}
   & \frac{1}{G_d} \to \frac{- L}{G_{d+1}\, (d-2)},
& & \alpha_1 \to - \frac{d}{4}\, \frac{c_d\, L^2}{(d-1) (d-2)},
& & \alpha_2 \to \frac{c_d\, L^2}{d-2},
& & \alpha_3 \to 0,
& & \bdyg_{\mu\nu} \to g_{\mu\nu},
& & \bdyD_\mu \to D_\mu.
\end{align*}


\section{The derivation of Eq.~(\ref{def:ren_charge_E})}
\label{Appendix_derivation}
According to the AdS/CFT dictionary~\cite{Maldacena:1997re}, the vacuum expectation value of the 
stress-energy tensor $\pExp{{\cal T}^{\mu\nu}}$ is derived from the variation of the 
bulk on-shell action~(\ref{bulk_action}) as
\begin{eqnarray}
\label{SE_tensor}
\pExp{{\cal T}^{\mu\nu}}
&=&\lim_{\varepsilon\to 0}\frac{2}{\sqrt{-\bdyg}}
\frac{\delta \Gamma}{\delta \bdyg_{\mu\nu}}\Biggl{|}_\text{On-shell}
=\lim_{\varepsilon\to 0}
\frac{1}{\Omega^{d+2}}\left[\frac{-K^{\mu\nu}+Kg^{\mu\nu}}{8\pi G_{d+1}}+2E_\text{ct}^{\mu\nu}   \right]. 
\end{eqnarray}

Let $n^M$ be the future-directed timelike unit normal vector to ${\cal S}_\varepsilon$ and $m^M$ be the outward spacelike unit normal vector 
to ${\cal S}_\varepsilon$.
Under the metric ansatz~(\ref{metric_ansatz_H}), these two unit normal vectors can be explicitly written as
\begin{eqnarray}
\label{normal_vec}
n^M=\Omega \sqrt{\frac{u}{f}}(1+\epsilon \xi T)^{-\frac{1}{2}}(\p_t)^M, \qquad 
m^M=-\Omega (\p_z)^M. 
\end{eqnarray}
Then, the bulk Noether charge~(\ref{bulk_NC_form}) can be expressed by the extrinsic curvature 
$K_{\mu\nu}$~(\ref{ext_curvature}) as 
\begin{eqnarray}
\label{Q_E}
 Q_\text{E} &=&\int_{{\cal S}_\varepsilon}{\bm Q}^{(d-1)}(\xi)
   =\int_{{\cal S}_\varepsilon}m^Nn^MQ_{MN} 
= \frac{1}{8\pi G_{d+1}}\int_{{\cal S}_\varepsilon}n_\mu {K^\mu}_\nu\xi^\nu
\end{eqnarray}
where in the second and third integrals, the volume element of ${\cal S}_\varepsilon^{(d-1)}$ is understood. 
Substituting Eq.~(\ref{Q_E}) into Eq.~(\ref{def:ren_charge_E}), we finally obtain 
\begin{eqnarray}
\label{Q_E_formula}
 Q^\text{(ren)}_\text{E}(\xi) &=& \lim_{\varepsilon\to 0}
   \left[
          \frac{1}{8\pi G_{d+1}}
                     \int_{{\cal S}_\varepsilon}n_\mu {K^\mu}_\nu\xi^\nu 
-\int_{{\cal S}_\varepsilon} \frac{K i_{\xi}{\bm \epsilon}^{(d)} }{8\pi G_{d+1}}
-2\int_{{\cal S}_\varepsilon} {\bm \epsilon}_{\mu\nu_2\nu_3\cdots \nu_d} 
  E_\text{ct}{}^\mu{}_\nu \xi^\nu\right]
\nonumber \\
&=& \lim_{\varepsilon\to 0}\int_{{\cal S}_\varepsilon} {\bm \epsilon}_{\mu\nu_2\nu_3\cdots \nu_d}
\left(\frac{{K^\mu}_\nu-{\delta^\mu}_\nu K}{8\pi G_{d+1}}-2{E_\text{ct}}^\mu{}_\nu\right)\xi^\nu
\nonumber \\
&=& -\int_{{\cal S}_\varepsilon} \tilde{\bm \epsilon}_{\mu\nu_2\nu_3\cdots \nu_d} 
\pExp{{{\cal T}^\mu}_\nu}\xi^\nu=-\int_{{\cal S}_\varepsilon}\tilde{n}_\mu 
\pExp{{{\cal T}^\mu}_\nu}\xi^\nu, 
\end{eqnarray}
where $\tilde{\bm \epsilon}^{(d)}=\lim_{\varepsilon \to 0}\Omega^d{\bm \epsilon}^{(d)}$ is the Levi-Civita volume form associated with the 
conformal metric $\bdyg_{\mu\nu}$, and in the last equality, we have used Eq.~(\ref{SE_tensor}). 
 
\section{Curvature tensors}
Following Ref.~\cite{Ishibashi:2025gvl}, the perturbations of the Ricci tensor $R^\mu_\nu$, higher curvature 
tensors $H^{(i)}_{\mu\nu}$ in Eqs.~(\ref{Eij_ct}) are derived as
\begin{subequations}
\label{H_pert_dS}
\begin{align}
\label{Ricci_bulk}
\delta (R_{\mu \sigma }g^{\nu \sigma}) =-\frac{\epsilon\Omega^2m^2}{2}\xi {H_\mu}^\nu, 
\end{align}
\begin{align}
\label{H1_pert_dS}
 \delta (H^{(1)}_{\mu \sigma} g^{\nu \sigma}) 
    =-\epsilon\frac{d(d-1)}{\ell^2}\Omega^4 m^2\xi {H_\mu}^\nu, 
 \end{align}
\begin{align}
\label{H2_pert_dS}
 \delta (H^{(2)}_{\mu \sigma}g^{\nu \sigma})=-\epsilon\Omega^4\frac{m^2}{2}\left(m^2+\frac{2(d-1)}{\ell^2} \right)\xi{H_\mu}^\nu. 
 \end{align}
 \end{subequations}



\begin{thebibliography}{99}

\bibitem{Bousso:2015mna}
R.~Bousso, Z.~Fisher, S.~Leichenauer and A.~C.~Wall,
``Quantum focusing conjecture,''
Phys. Rev. D \textbf{93} (2016) no.6, 064044
doi:10.1103/PhysRevD.93.064044
[arXiv:1506.02669 [hep-th]].
\bibitem{Penrose}
R.~Penrose, Gravitational collapse and space-time singularities, 
Phys.~Rev.~Lett. {\bf 14}, 57~(1967).  

\bibitem{Wall:2010jtc}
A.~C.~Wall,
``The Generalized Second Law implies a Quantum Singularity Theorem,''
Class. Quant. Grav. \textbf{30} (2013), 165003
[erratum: Class. Quant. Grav. \textbf{30} (2013), 199501]
doi:10.1088/0264-9381/30/19/199501
[arXiv:1010.5513 [gr-qc]].

\bibitem{Bousso:2025xyc}
R.~Bousso,
``Robust Singularity Theorem,''
Phys. Rev. Lett. \textbf{135} (2025) no.1, 011501
doi:10.1103/6f9b-3jmx
[arXiv:2501.17910 [hep-th]].

\bibitem{Engelhardt:2026qqs}
N.~Engelhardt and I.~Nagar,
``A Quantum Singularity Theorem for the Evaporating Black Hole,''
[arXiv:2605.05326 [hep-th]].

\bibitem{Engelhardt:2014gca}
N.~Engelhardt and A.~C.~Wall,
``Quantum Extremal Surfaces: Holographic Entanglement Entropy beyond the Classical Regime,''
JHEP \textbf{01} (2015), 073
doi:10.1007/JHEP01(2015)073
[arXiv:1408.3203 [hep-th]].

\bibitem{Penington:2019npb}
G.~Penington,
``Entanglement Wedge Reconstruction and the Information Paradox,''
JHEP \textbf{09} (2020), 002
doi:10.1007/JHEP09(2020)002
[arXiv:1905.08255 [hep-th]].

\bibitem{Page:1993wv}
D.~N.~Page,
``Information in black hole radiation,''
Phys. Rev. Lett. \textbf{71} (1993), 3743-3746
doi:10.1103/PhysRevLett.71.3743
[arXiv:hep-th/9306083 [hep-th]].

\bibitem{Iizuka:2025xnd}
N.~Iizuka, A.~Ishibashi, K.~Maeda, H.~Nakayama and T.~Nishioka,
Can. J. Phys. \textbf{104}, 1-33 (2026)
doi:10.1139/cjp-2025-0271
[arXiv:2509.01286 [hep-th]].


\bibitem{Ryu:2006bv}
S.~Ryu and T.~Takayanagi,
``Holographic derivation of entanglement entropy from AdS/CFT,''
Phys. Rev. Lett. \textbf{96} (2006), 181602
doi:10.1103/PhysRevLett.96.181602
[arXiv:hep-th/0603001 [hep-th]].

\bibitem{Maldacena:1997re}
J.~M.~Maldacena,
``The Large N limit of superconformal field theories and supergravity,''
Adv. Theor. Math. Phys. \textbf{2} (1998), 231-252
doi:10.4310/ATMP.1998.v2.n2.a1
[arXiv:hep-th/9711200 [hep-th]].

\cite{Wald:1993nt}
\bibitem{Wald:1993nt}
R.~M.~Wald,
Phys. Rev. D \textbf{48} (1993) no.8, R3427-R3431
doi:10.1103/PhysRevD.48.R3427
[arXiv:gr-qc/9307038 [gr-qc]].
\bibitem{Faulkner:2013ica}
T.~Faulkner, M.~Guica, T.~Hartman, R.~C.~Myers and M.~Van Raamsdonk,
``Gravitation from Entanglement in Holographic CFTs,''
JHEP \textbf{03} (2014), 051
doi:10.1007/JHEP03(2014)051
[arXiv:1312.7856 [hep-th]].

\bibitem{Ishibashi:2021fcc}
A.~Ishibashi and K.~Maeda,
Phys. Rev. D \textbf{104} (2021) no.2, 026004
doi:10.1103/PhysRevD.104.026004
[arXiv:2104.01862 [hep-th]].
\bibitem{Iyer:1994ys}
V.~Iyer and R.~M.~Wald,
``Some properties of Noether charge and a proposal for dynamical black hole entropy,''
Phys. Rev. D \textbf{50} (1994), 846-864
doi:10.1103/PhysRevD.50.846
[arXiv:gr-qc/9403028 [gr-qc]].

\bibitem{Compere:2008us}
  G.~Compere and D.~Marolf,
  ``Setting the boundary free in AdS/CFT,''
  Class. Quant. Grav. \textbf{25}, 195014 (2008)
 doi:10.1088/0264-9381/25/19/195014
  [arXiv:0805.1902 [hep-th]].
\bibitem{Ishibashi:2023luz}
A.~Ishibashi, K.~Maeda and T.~Okamura,
``Semiclassical Einstein equations from holography and boundary dynamics,''
JHEP \textbf{05} (2023), 212
doi:10.1007/JHEP05(2023)212
[arXiv:2301.12170 [hep-th]].

\bibitem{Maldacena:2012xp}
J.~Maldacena and G.~L.~Pimentel,
JHEP \textbf{02} (2013), 038
doi:10.1007/JHEP02(2013)038
[arXiv:1210.7244 [hep-th]].

\bibitem{Narayan:2015vda}
K.~Narayan,
Phys. Rev. D \textbf{91} (2015) no.12, 126011
doi:10.1103/PhysRevD.91.126011
[arXiv:1501.03019 [hep-th]].

\bibitem{Gibbons:1977mu}
G.~W.~Gibbons and S.~W.~Hawking,
Phys. Rev. D \textbf{15} (1977), 2738-2751
doi:10.1103/PhysRevD.15.2738
\cite{Chandrasekaran:2022cip}
\bibitem{Chandrasekaran:2022cip}
V.~Chandrasekaran, R.~Longo, G.~Penington and E.~Witten,
JHEP \textbf{02} (2023), 082
doi:10.1007/JHEP02(2023)082
[arXiv:2206.10780 [hep-th]].

\bibitem{Ishibashi:2025gvl}
A.~Ishibashi, K.~Maeda and T.~Okamura,
JHEP \textbf{05} (2026), 308
doi:10.1007/JHEP05(2026)308
[arXiv:2512.00503 [hep-th]].

\bibitem{Ishibashi:2024fnm}
A.~Ishibashi, K.~Maeda and T.~Okamura,
JHEP \textbf{04} (2025), 167
doi:10.1007/JHEP04(2025)167
[arXiv:2412.18764 [hep-th]].

\bibitem{Bunting:2015sfa}
W.~Bunting, Z.~Fu and D.~Marolf,
Class. Quant. Grav. \textbf{33} (2016) no.5, 055008
doi:10.1088/0264-9381/33/5/055008
[arXiv:1509.00074 [hep-th]].

\bibitem{HawkingEllis}
S.~W.~Hawking and G.~F.~R.~Ellis,
\textit{The Large Scale Structure of Space-Time}
(Cambridge University Press, Cambridge, 1973).
\end{thebibliography}
\end{document}